\documentclass[preprint,prb]{revtex4-1}

\usepackage{hyperref}
\usepackage{amsmath}
\usepackage{array}
\usepackage{dcolumn}
\usepackage{graphicx}
\usepackage{siunitx}

\begin{document}

\title{{\em Ab-initio} free energies of liquid metal alloys: application to the phase diagrams of Li-Na and Na-K}

\author{Yang Huang and Michael Widom}
\affiliation{Department of Physics, Carnegie Mellon University, Pittsburgh, PA  15213}

\author{Michael C. Gao}
\affiliation{National Energy Technology Laboratory, Albany OR 97321}

\begin{abstract}
  Comparison of free energies between different phases and different compositions underlies the prediction of alloy phase diagrams. To allow direct comparison, consistent reference points for the energies or enthalpies are required, and the entropy must be placed on an absolute scale, yielding {\em absolute} free energies.  Here we derive absolute free energies of liquids from {\em ab-initio} molecular dynamics (AIMD) by combining the directly simulated enthalpies with an entropy derived from simulated densities and pair correlation functions. As an example of the power of this method we calculate the phase diagrams of two binary alkali metal alloys, Li-Na and K-Na, revealing a critical point and liquid-liquid phase separation in the former case, and a deep eutectic in the latter. Good agreement with experimental data demonstrates the power of this simple method.
\end{abstract}
\maketitle{}

\section{Introduction}
\label{sec:Intro}

The free energy of a compound, $G=H-TS$, depends on both the enthalpy $H$ and the entropy $S$. Shifting $H$ by a constant value has no consequence because only free energy {\em differences} enter into thermodynamics. Although the values are arbitrary, comparing the free energies of competing phases is facilitated if consistently chosen reference values are applied to the enthalpies. By thermodynamic convention, the reference point for enthalpy of a compound is chosen as the sum of the enthalpies of all constituent elements in their stable state at standard temperature and pressure. However, other choices can be equally valid, in principle. For example, one could choose the enthalpy as calculated within density functional theory at temperature $T=0K$ as the reference point, placing high temperature enthalpies as calculated from {\em ab-initio} molecular dynamics (AIMD) on a well-defined absolute scale. In contrast, since the entropy enters the free energy multiplied by $T$, its actual value cannot be chosen arbitrarily. Indeed, a unique reference point for the entropy is provided by the Third Law, namely $S$ must vanish at $T=0K$. The combination of enthalpy on a well-defined relative scale and entropy on an absolute scale yields absolute free energy. 

Knowledge of absolute entropy $S(E)$ is equivalent to knowledge of the configurational density of states $\Omega(E)$. Histogram~\cite{Ferrenberg89} and entropic sampling methods~\cite{entropic} such as Wang-Landau~\cite{Wang2001} calculate $\Omega(E)$ up to an unknown constant factor, yielding relative but not absolute entropies. In some special discrete cases, such as lattice models where the total number of states is known, $\Omega(E)$ can be normalized yielding absolute entropy.

Most methods to compute absolute free energy rely on connecting the free energy of interest to some reference state of known free energy. Thermodynamic integration~\cite{Kirkwood1935} calculates $\partial G/\partial \lambda$ through simulation ($\lambda$ is some parameter in the Hamiltonian) then numerically integrates this derivative. Thermodynamic perturbation theory~\cite{Zwanzig54,Peter2004} expresses $G(\lambda)$ as a low-order Taylor series expansion. Umbrella sampling~\cite{Torrie1977} and the Bennett acceptance ratio method~\cite{Bennett76} provide increased computational efficiency to these basic approaches, as do other schemes~\cite{Grabowski17}. In practice, the thermodynamic integration requires reversible paths so that the free energy and its derivatives are well-defined. This is sometimes referred to as the slow growth approach\cite{Woo1996,Woo1997,Hu2002}. Jarzynski's identity $exp(-\beta \Delta F) = <exp(-\beta W_\lambda)>$ holds even for nonequilibrium transitions and allows for evolution over short time durations, which is known as the "fast growth" method~\cite{Hendrix2001}.

Other methods build the free energy through sequential addition of particles. The exact scanning approach \cite{Meirovitch1982,Meirovitch1988,Meirovitch2009} computes the partial density of states $\rho(\alpha_k|\alpha_{k-1}\cdots\alpha_1)$ where the set $\{\alpha_j\}$ is an ordered sequence of states containing successively more particles. Similarly, the particle insertion method\cite{Widom1963} calculates the free energy difference of a $k$-atom system and a $(k-1)$-atom system ({\em i.e.} the chemical potential), although without explicitly calculating the density of states.

The empirical CALPHAD approach~\cite{CALPHAD,alma991016488169704436}, proposes analytical free energy models for the Gibbs free energy $G(x,T)$ of a compound with composition $x$. Starting from the ideal free energy $G_{\rm ideal}=H_{\rm ideal}-TS_{\rm ideal}$, CALPHAD models the excess free energy in a series of Redlich-Kister polynomials~\cite{RedlichKister} with coefficients obtained from experimental information such as heat capacity and phase diagrams. The result is a set of free energy functions in analytic form that can be used to interpolate the free energy into compositions for which no data is available.

We recently developed an approximate method to calculate absolute entropy of liquid metals from AIMD simulations of their densities and pair correlation functions~\cite{Gao2018,Widom2019}. Since entropies cannot be derived directly from simulations, our method provides a feasible approach to calculate the absolute entropy, and hence the absolute free energy, with high accuracy and reduced computational effort. As a demonstration of the utility of absolute free energy and the power of our calculational approach, we apply the method to calculate the free energies of two binary alkali metal alloy systems, Li-Na and K-Na. These examples are chosen because, despite the seeming chemical similarity, the two systems exhibit very different phase behaviors. Li and Na are nearly immiscible, both in the solid state and in the liquid below a critical point at $T=578K$. In contrast, K-Na compounds remain liquid below the melting points of elemental Na and K, forming a deep eutectic at $33$\% Na and $T=260K$ ($-13C$). The low melting point of K$_2$Na (often abbreviated as ``NaK'') makes the liquid alloy useful as a coolant for nuclear reactors~\cite{Chetal2001} and other applications. The immiscibility of Na in Li makes it potentially useful for supression of dendrites in Li-ion batteries~\cite{Stark2011}.

In the following we first describe our simulation methods, including the manner in which we obtain absolute enthalpies and entropies. Then we validate the methods by comparing our calculated densities for pure elemental liquids with experimental values, and our absolute entropies with values tabulated in the NIST-JANAF tables~\cite{JANAF}. Finally, we present our predicted phase diagrams of Li-Na and K-Na and compare with published experimental results and they show a good agreement. We find a high positive energy of mixing between Li and Na atoms which leads to phase separation at moderately high temperatures, with a critical point for phase mixing at higher temperatures. For the K-Na system, the energy of mixing is still positive, but it is relatively weak. In consequence the entropy dominates the free energy, and after incorporating the Gibbs free energies of competing solid phases we observe a deep eutectic transition at temperatures below 0C.

\section{Methods}
\label{sec:Methods}
We simulate liquid K-Na and Li-Na using {\em ab-initio} energies and forces to accurately reproduce their configurational ensembles. Different strategies are used for elemental and binary metallic systems. For pure elements standard {\em ab-initio} molecular dynamics (AIMD) simulations are performed. For binary alloys,we supplement AIMD with additional Monte Carlo chemical species swapping steps\cite{Widom2014} in order to accelerate the sampling of diverse configurations. Enthalpies are taken directly from the {\em ab-initio} total energies, while entropies are obtained from integrals of correlation functions~\cite{Gao2018,Widom2019}. We carry out our simulations in canonical ensembles although our entropy model is expressed in the grand canonical ensemble, relying on locality of the correlations to achieve ensemble independence~\cite{Evans1989}.  

Specific simulated temperatures and compositions are chosen to cover the relevant soluble regions of the Li-Na and K-Na phase diagrams. Only soluble phases are chosen to avoid contaminated our data with multiple phases and interfacial free energies. We then fit the Gibbs free energy to an analytical model from which we derive the phase diagrams by computing the convex hull of $G(x,T)$. Individual data points are given in the Supplemental Material.

\subsection{\em Ab-initio molecular dynamics and Monte Carlo}
\label{sec:AIMD}
Our AIMD simulations apply electronic density  functional theory as implemented in the Vienna Ab-initio Simulation Package ({\tt VASP}~\cite{Kresse1996, Kresse1999}). First-principles energies and forces are calculated using the PBE generalized gradient approximation \cite{Perdew1996,Blochl1994}. MD time steps are set at 1fs with the temperature controlled in the $NVT$ ensemble using a Nose thermostat. We take a plane-wave-basis set with a cut-off energy of $E=300$eV. Semi-core electrons are included in the pseudopotentials for potassium atoms and sodium atoms while only valence electrons are considered for lithium atoms. We employ simulation cells of 300 atoms for K-Na and 500 atoms for Li-Na. Justification for these decisions is presented in Appendix A. 

Simulations at a given temperature and composition are pre-annealed for a minimum of 1 ps, until the onset of equilibrium energy fluctuations, followed by data acquisition for a minimum of 2 ps. We take equilibrated configurations from high temperature runs as initial conditions for lower temperature runs. To predict the density at a given temperature and composition, we monitor the total pressure at five different volumes, and then find the volume at which the pressure vanishes by fitting to a quadratic.

In binary systems, we additionally perform Metropolis Monte Carlo by testing a randomly chosen interchange of two atoms' chemical species and accepting the change with probability $\exp{(-\Delta E/k_B T)}$. We attempt one species swap every 10 MD steps. On average, a total number of $300$ atomic swap attempts are made with a acceptance rates around $15\%$. Supplementing AIMD with Monte Carlo (MCMD~\cite{Widom2014}) accelerates the approach to equilibrium and enhances the configurational diversity of the simulated ensemble, as discussed in Appendix B.

\subsection{Entropy}
\label{sec:S}
We calculate absolute entropies directly from MCMD simulations performed at the temperatures, densities and compositions of interest by evaluating the leading terms in an expansion of the entropy in a series of progressively higher-order correlation functions~\cite{Green1952,Raveche1971a,Wallace1987,Evans1989}. This method has been previously validated for elemental liquid Al and Cu, and applied to the AlCu binary liquid alloy~\cite{Gao2018,Widom2019}.

The quantum mechanically derived absolute entropy of the ideal gas is
\begin{equation}
  \label{eq:s1}
  S_{\rm Ideal}/k_B = \frac{5}{2} - \sum_\alpha x_\alpha \log(\rho x_\alpha \lambda_\alpha ^3),
\end{equation}
where $\alpha$ denotes atomic species, $\lambda_\alpha = \sqrt{h^2/2\pi m_{\alpha} k_{\rm B} T}$ is the thermal de Broglie wavelength of species $\alpha$, $x_\alpha$ is its fraction, and $\rho$ is total atomic number density. Note that $S_{\rm Ideal}$ contains the entropy of mixing
\begin{equation}
  \label{eq:Smix}
  S_{\rm Mix}/k_B = -\sum_\alpha x_\alpha\log(x_\alpha).
\end{equation}
The leading term in the correlation function expansion is the single-body entropy $S_1=S_{\rm Ideal}-k_B$, with the difference arising from interchange of multiple atoms~\cite{Evans1989,Widom2019}. The two-body corrections to $S_1$ include a fluctuation term
\begin{equation}
  \label{eq:sfluct}
  S_{\rm fluct} = \frac{1}{2} \sum_{\alpha,\beta}x_\alpha x_\beta \left(1 + \rho \int dr\, 4\pi r^2~ (g_{\alpha\beta}(r)-1)\right)
\end{equation}
that is positive but very small (it is proportional to the isothermal compressibility) and an information term
\begin{equation}
  \label{eq:sinfo}
  S_{\rm info} = -\frac{1}{2}\rho \sum_{\alpha,\beta}x_\alpha x_\beta \int dr\, 4\pi r^2~ g_{\alpha\beta}(r)\ln g_{\alpha\beta}(r)
\end{equation}
that is negative-definite and reflects the entropy reduction due to the information content of the pair correlation functions. We approximate the total entropy as $S\approx S_1+S_2$ with $S_2=S_{\rm fluct}+S_{\rm info}$. Note that $S_2$ can be decomposed into partial contributions for each species pair, $S_2=\sum_{\alpha\beta}x_{\alpha}x_{\beta}S_{\alpha\beta}$.

\begin{figure}{}
  \label{fig:g_s2}
  \includegraphics[width=.49\textwidth, height=.3\textwidth]{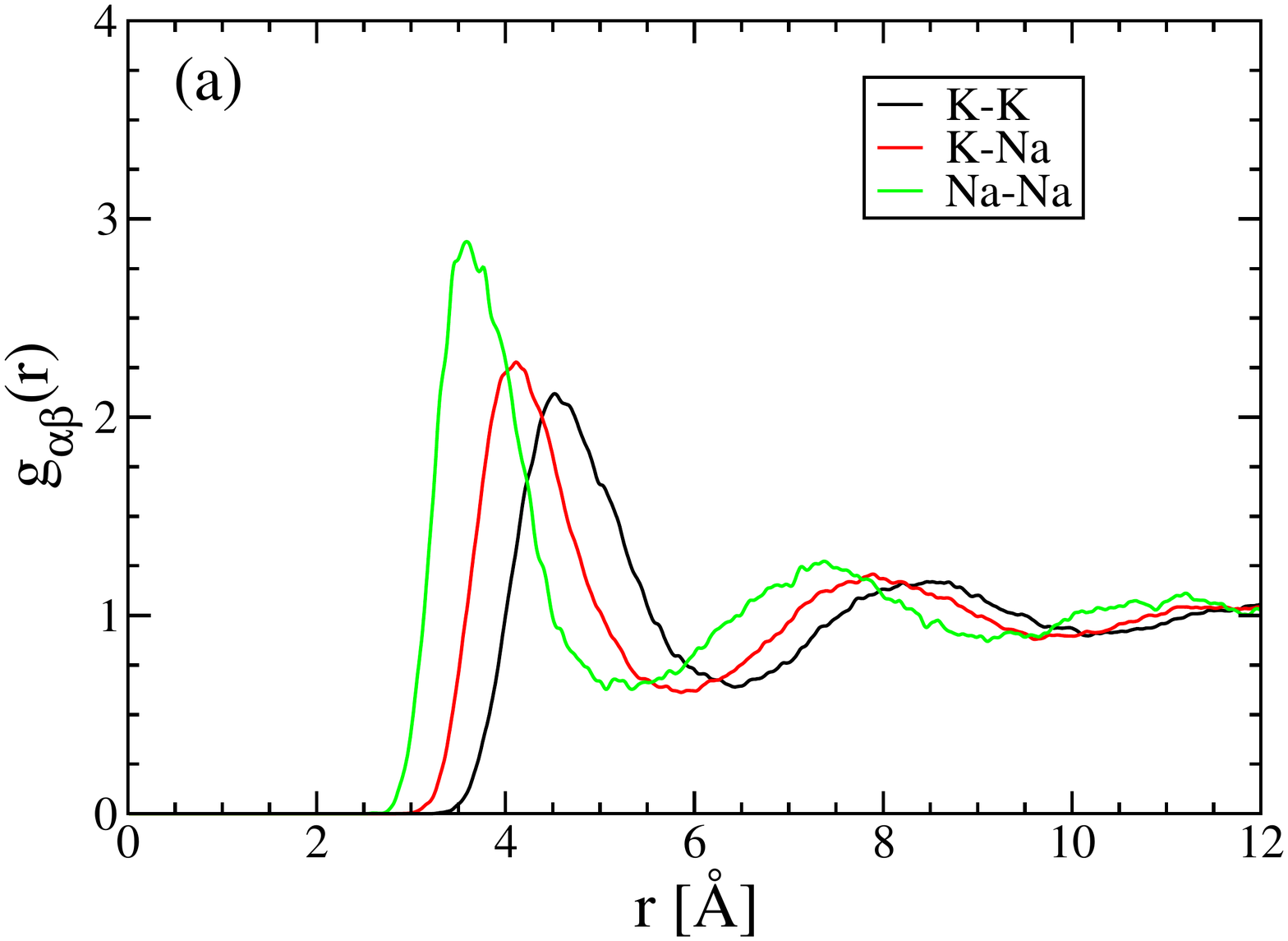}
  \includegraphics[width=.49\textwidth, height=.3\textwidth]{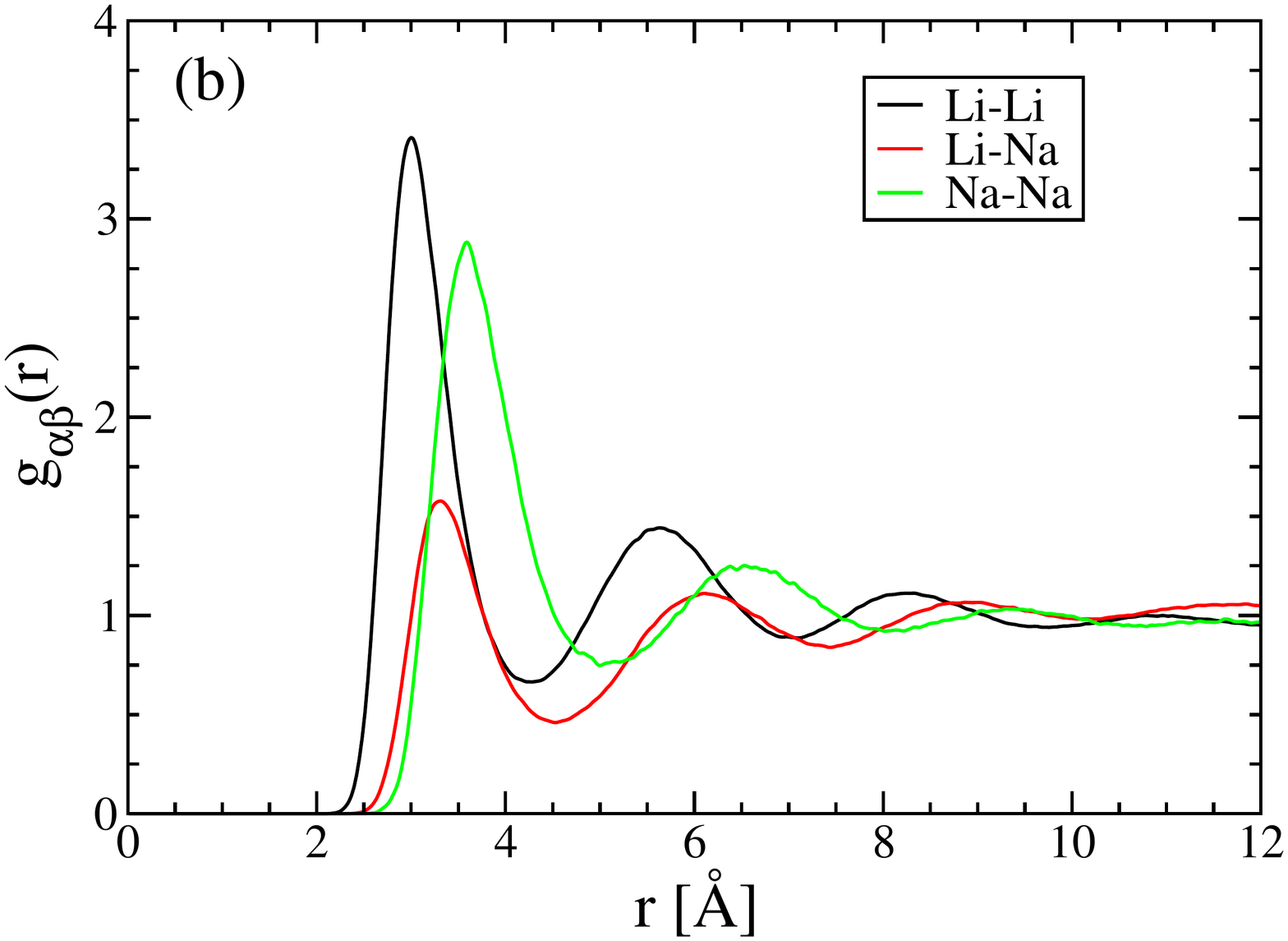}
  \includegraphics[width=.49\textwidth, height=.3\textwidth]{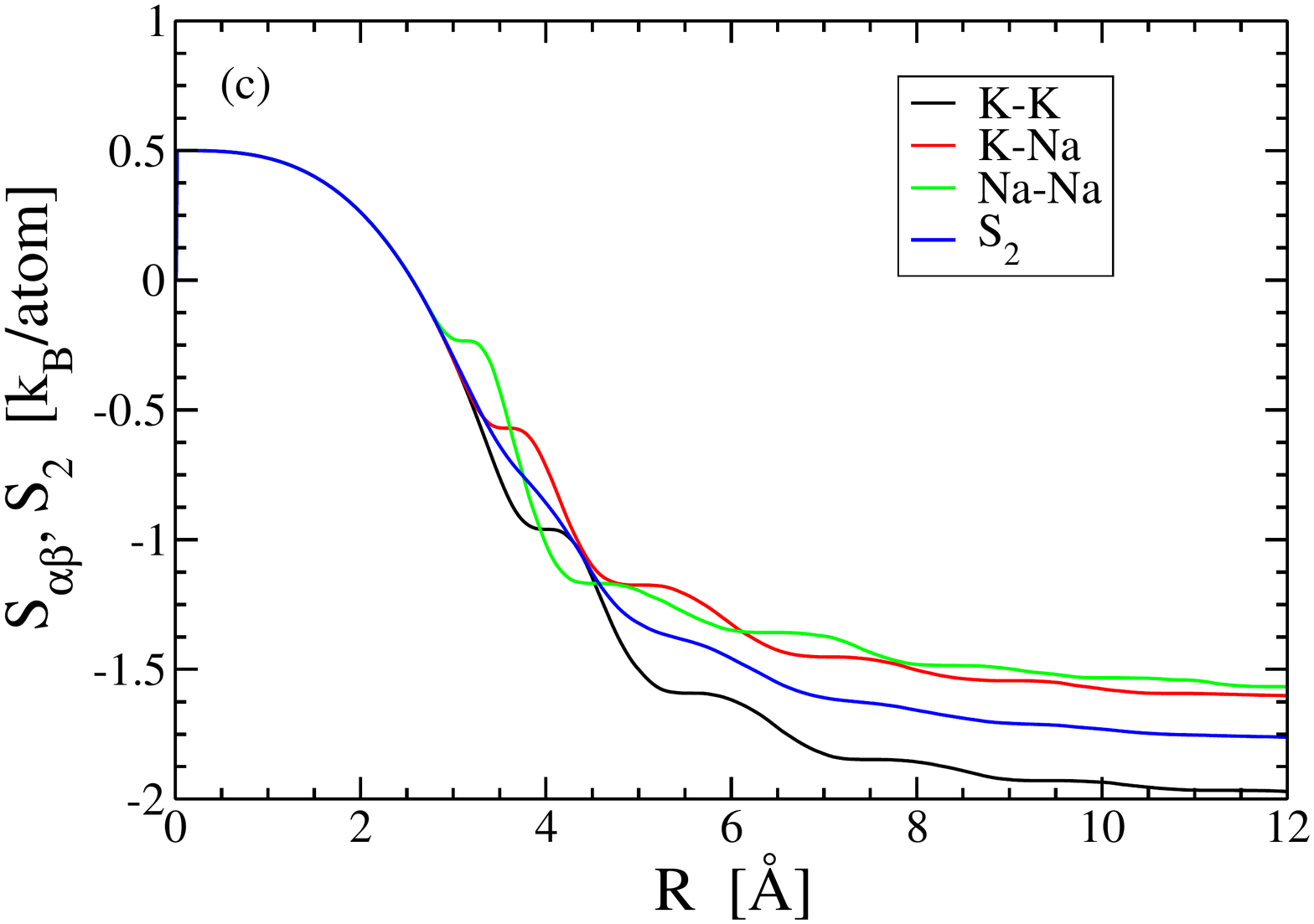}
  \includegraphics[width=.49\textwidth, height=.3\textwidth]{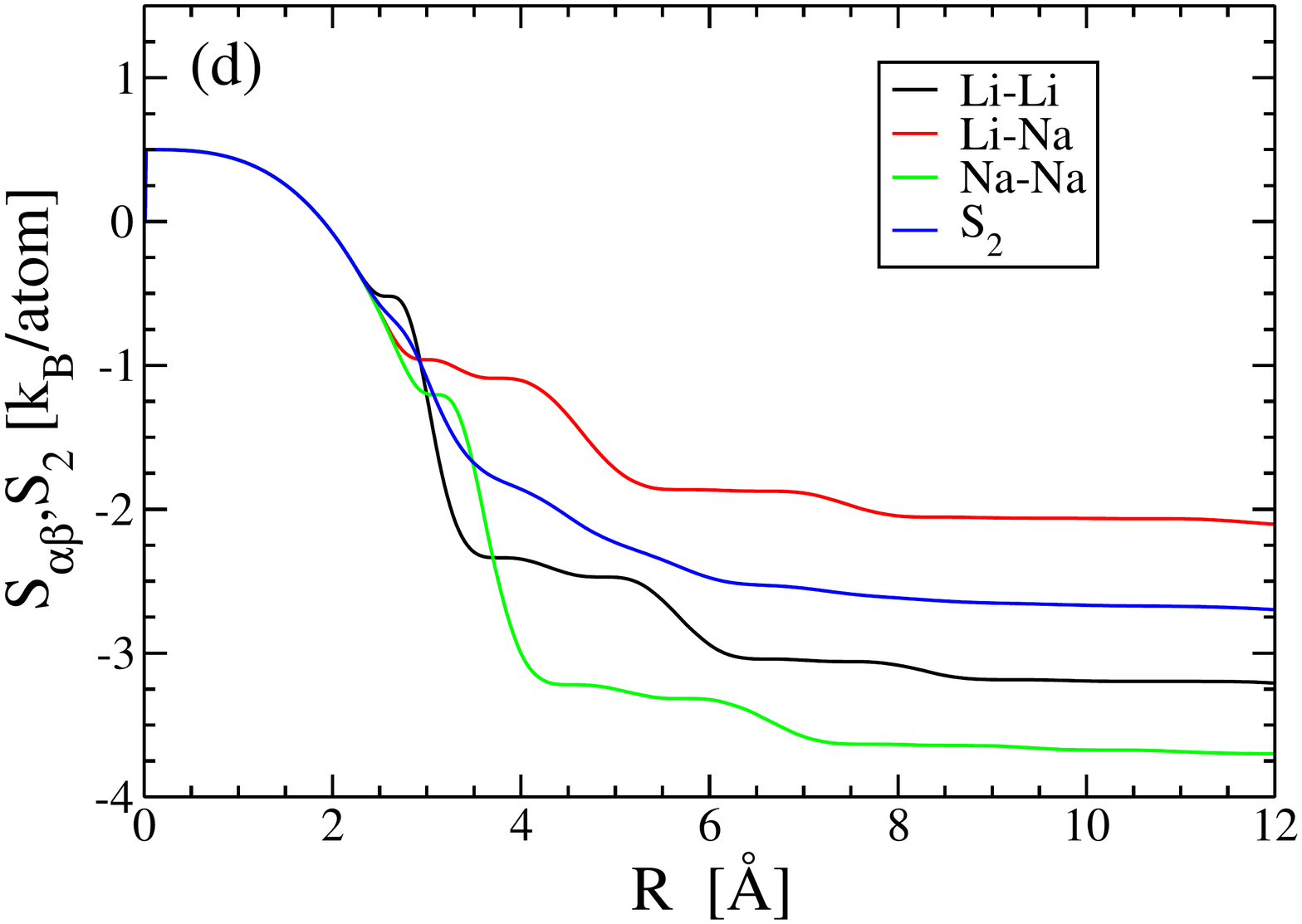}
  \caption{Radial distribution functions $g_{\alpha\beta}(r)$ of (a) K$_2$Na and (b) Li$_2$Na for 300 atoms at T=$473$K. (c) and (d) Two-body entropies $S_{\alpha\beta}$ and $S_2$ integrated up to distance $R$.}
\end{figure}

Figs.~\ref{fig:g_s2}~(a) (c) show correlation functions of K$_2$Na and Li$_2$Na, respectively. Each correlation function vanishes within its atomic core, and thereafter exhibits decaying oscillations. Positions of the first peaks vary in accordance with relative atomic diameters. The oscillation frequencies are similar for each combination of species, suggesting a universal origin of oscillation. Indeed it is known that alkali metals possess long-range oscillatory potentials\cite{Arthur1966} with a frequency of twice the Fermi wavenumber\cite{Hafner1986}. For valence-1 elements with atomic volume $v=1/\rho$, the Fermi wavenumbers $k_F=(3\pi^2/v)^{1/3}$ evaluate to $k_F=0.75$ and $1.01$\AA$^{-1}$ for K$_2$Na and Li$_2$Na, respectively. These values roughly match the $k_F$ values $0.84$ and $1.04$\AA$^{-1}$ inferred from observed oscillation frequencies of the correlations. A hard sphere distribution with hard sphere radius $R=(3v/4\pi)^{1/3}$ also matches the frequencies but does not reproduce the detailed shapes of the correlation functions.

Because the oscillations decay, the integrals in Eqs.~(\ref{eq:sfluct}) and~(\ref{eq:sinfo}) converge as the upper limit of integration $R$ increases, as shown in Figs.~\ref{fig:g_s2}(b) and (d). We take the values at $R=12$~\AA for our values of $S_2$.  $S_{\alpha\beta}$ and $S_2$ are negative definite (in the large $R$ limit) reflecting the loss of entropy due to the correlations. The entropy loss is larger for Li$_2$Na than for K$_2$Na because of the strong chemical order that prefers like neighbors (Li-Li and Na-Na), while this effect is nearly absent in the case of K$_2$Na.

\subsection{Electronic free energy}
Electronic free energies are included in every case to supplement our systematic calculations of thermodynamic quantities of alkali metals. Electronic free energies at finite temperature are obtained from the DFT-predicted electronic density of state $D(\epsilon)$ as discussed in~\cite{WidomJMR2018}. Electronic contributions to the relative free energies $\Delta G$ are relatively small compared with $\Delta G$ itself.

\subsection{Interpolation}
Because we carry out simulations at discrete temperatures and compositions, but we wish to determine phase boundaries as continuously varying functions of temperature, we require a method to interpolate the enthalpy and entropy. To compute the phase diagrams it suffices to model  $\Delta H(x,T)$ and $\Delta S(x,T)$ relative to their values at the concentration endpoints $x=0$ and $x=1$. Then the compositions $x$ where $\Delta G(x,T)=\Delta H(x,T)-T\Delta S(x,T)$ lies above its own convex hull determine the phase coexistence regions.

We fit the excess enthalpy $\Delta H(x,T) $ and two-body term $S_2(x,T)$ to a quartic polynomial
\begin{equation}
  \label{eq:fit}
  f(x,T)=x(1-x)\left[a(T)x+b(T)x+C(T)x^2\right].
\end{equation}
where $a(T)$, $b(T)$ and $c(T)$ are linear functions of $T$, resulting in 6 fitting parameters for each thermodynamic function $f$. The enthalpy and entropy satisfy the constraints $f(0)=0$ and $f(1)=0$, while the quadratic function of composition in the bracket captures asymmetry The simple linear temperature dependence approximation is designed for accuracy over a narrow temperature range. This approximation works well for single species liquid enthalpies as shown in Fig.~\ref{fig:result-entropy}.  The $S_1$ term is calculated from Eq.~\ref{eq:s1} using a quartic function to fit the composition-dependent density. By this approach, we capture the logarithmic singularities of the entropy near $x=0$ and $x=1$.

\subsection{Solid phases}
\label{sec:Solids}
The Gibbs free energies of competing phases must be included to determine the global phase diagrams. These phases are body centered cubic solid phases of pure elements and a Laves phase (Pearson hP12, Strukturbericht C14) KNa$_2$ binary phase. The Gibbs free energy for a solid phase includes the vibrational free energy $G^{v}$, the electronic free energy $G^e$, and a configurational free energy $G^{c}$. In principle the configurational term includes contributions due to chemical and vacancy disorder~\cite{WidomJMR2018}, however experimental evidence suggests that the K-Li-Na solid phases are nearly stoichiometric, so we simply approximate $G^c$ with the enthalpy of the fully relaxed ({\em i.e.} $T$=0K) structure. First-principle vibrational free energy calculations use the same pseudopotentials and exchange-correlation function as for the liquid simulations, but with an increased plane wave cut-off energy of 500eV and an increased electronic $k$-mesh density so that we may obtain accurate interatomic force constants. The Gibbs free energies at finite temperatures are calculated within the quasi-harmonic approximation using {\tt Phonopy}~\cite{phonopy, phonopy-qha}. The differing cutoff energies and $k$-meshes result in an offset between solid and liquid enthalpies, which we correct by matching our calculated enthalpy differences between 200K and 500K to experiment.

\section{Results}
\label{sec:Results}

\subsection{Pure elemental Li, Na and K}
\label{sec:elements}
\begin{figure*}[htpb]
  \centering
  \includegraphics[width=.3\textwidth, height=.2\textwidth]{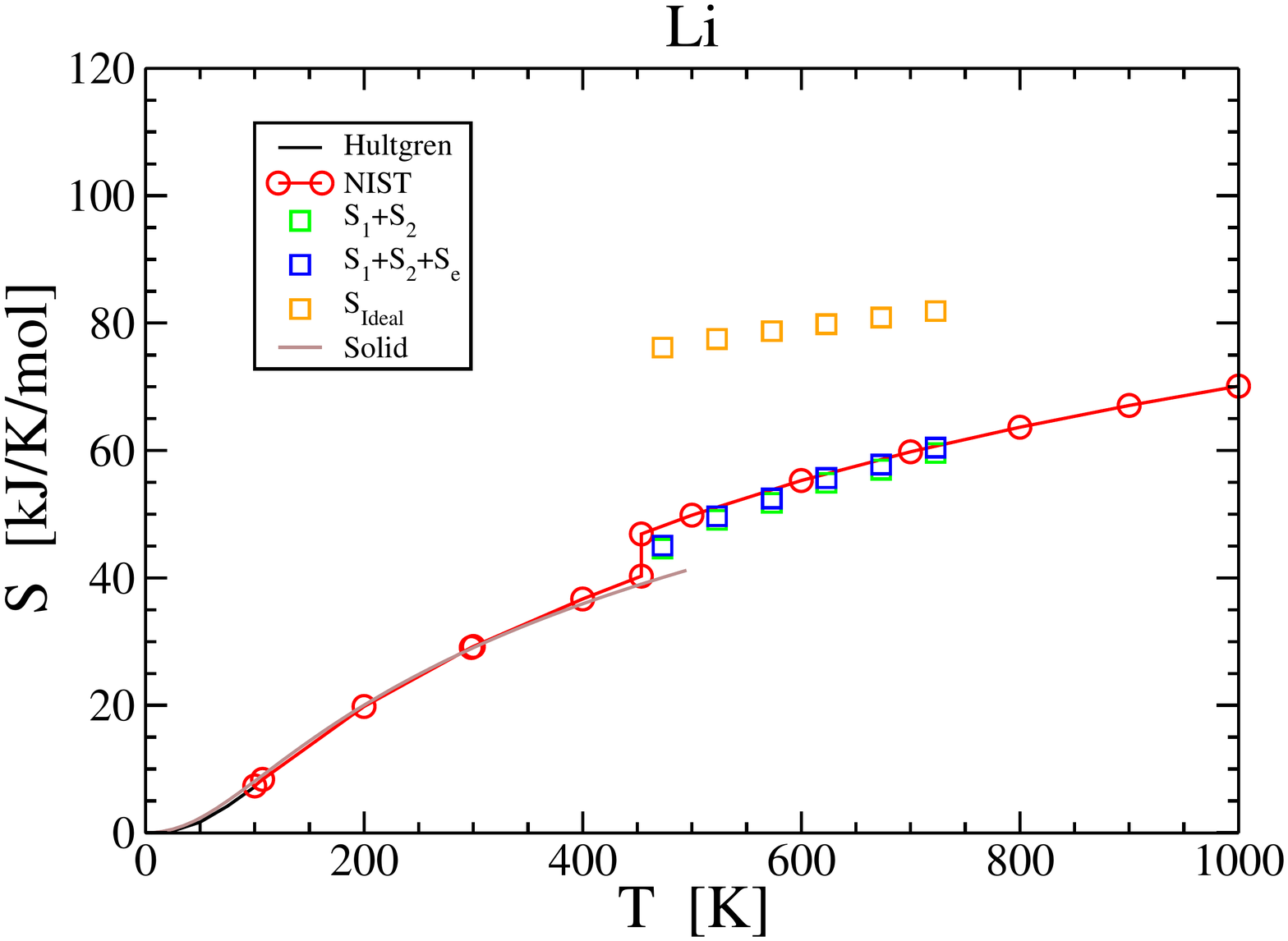}
  \includegraphics[width=.3\textwidth, height=.2\textwidth]{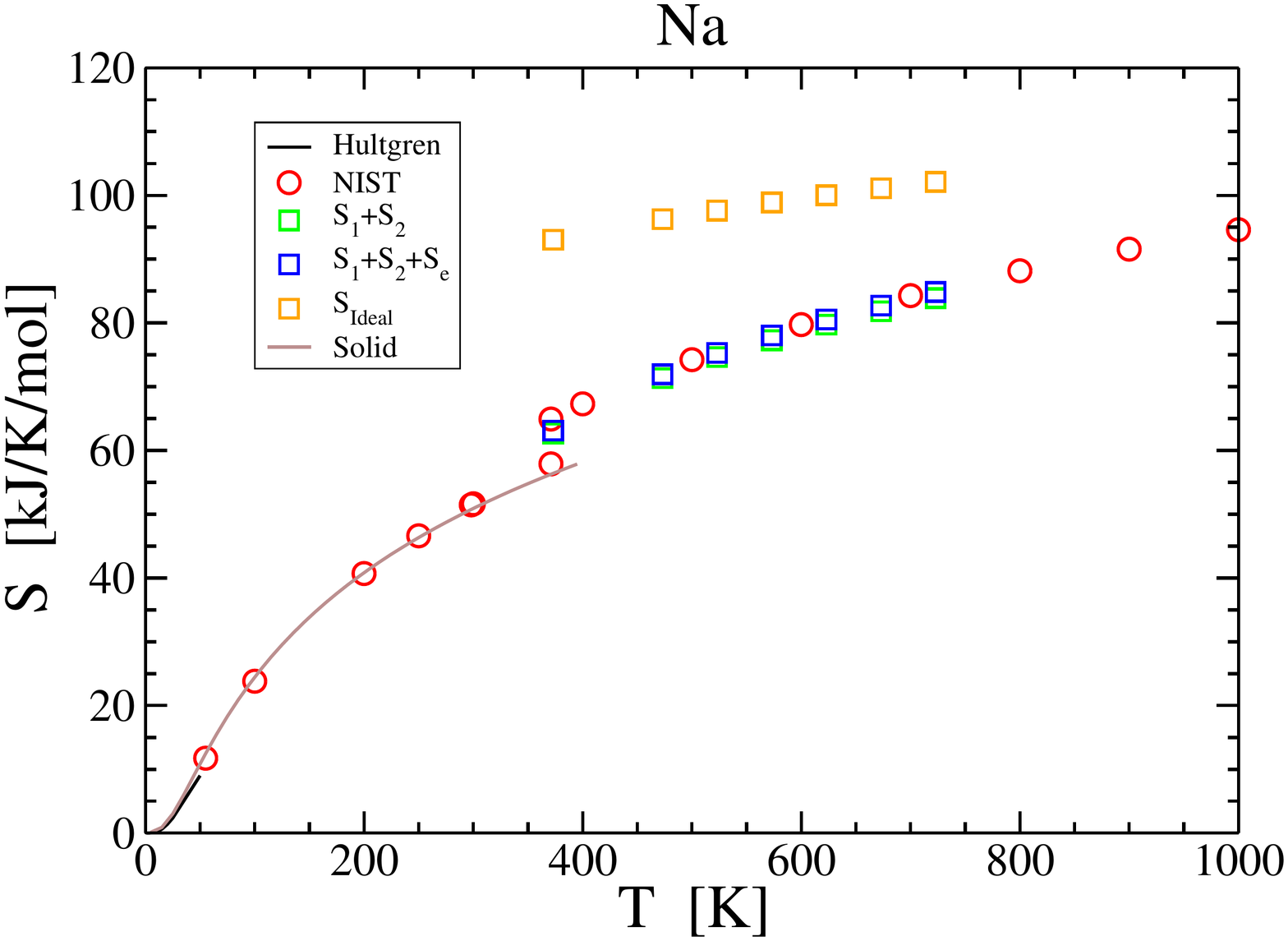}
  \includegraphics[width=.3\textwidth, height=.2\textwidth]{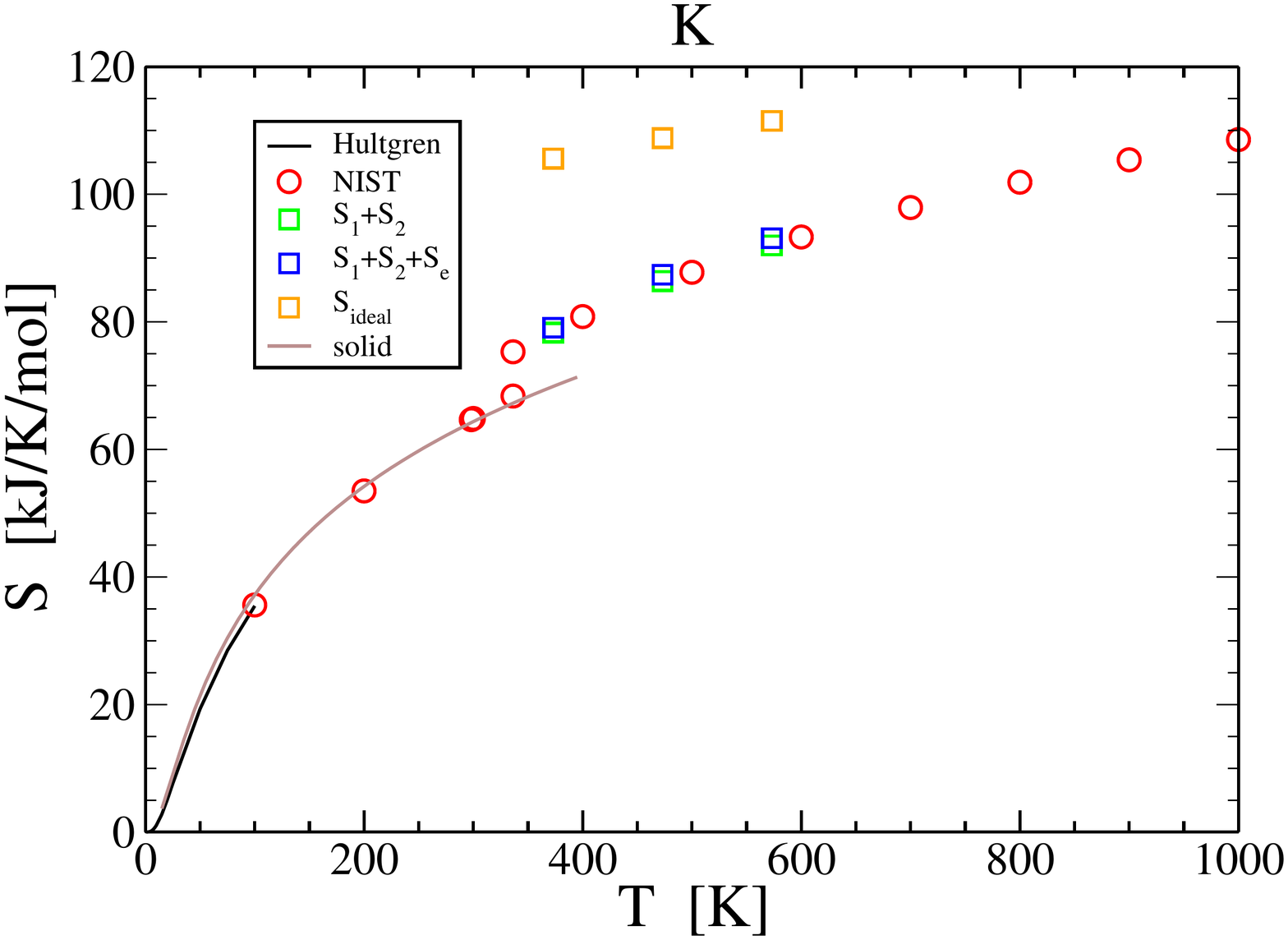}\\
  \includegraphics[width=.3\textwidth, height=.2\textwidth]{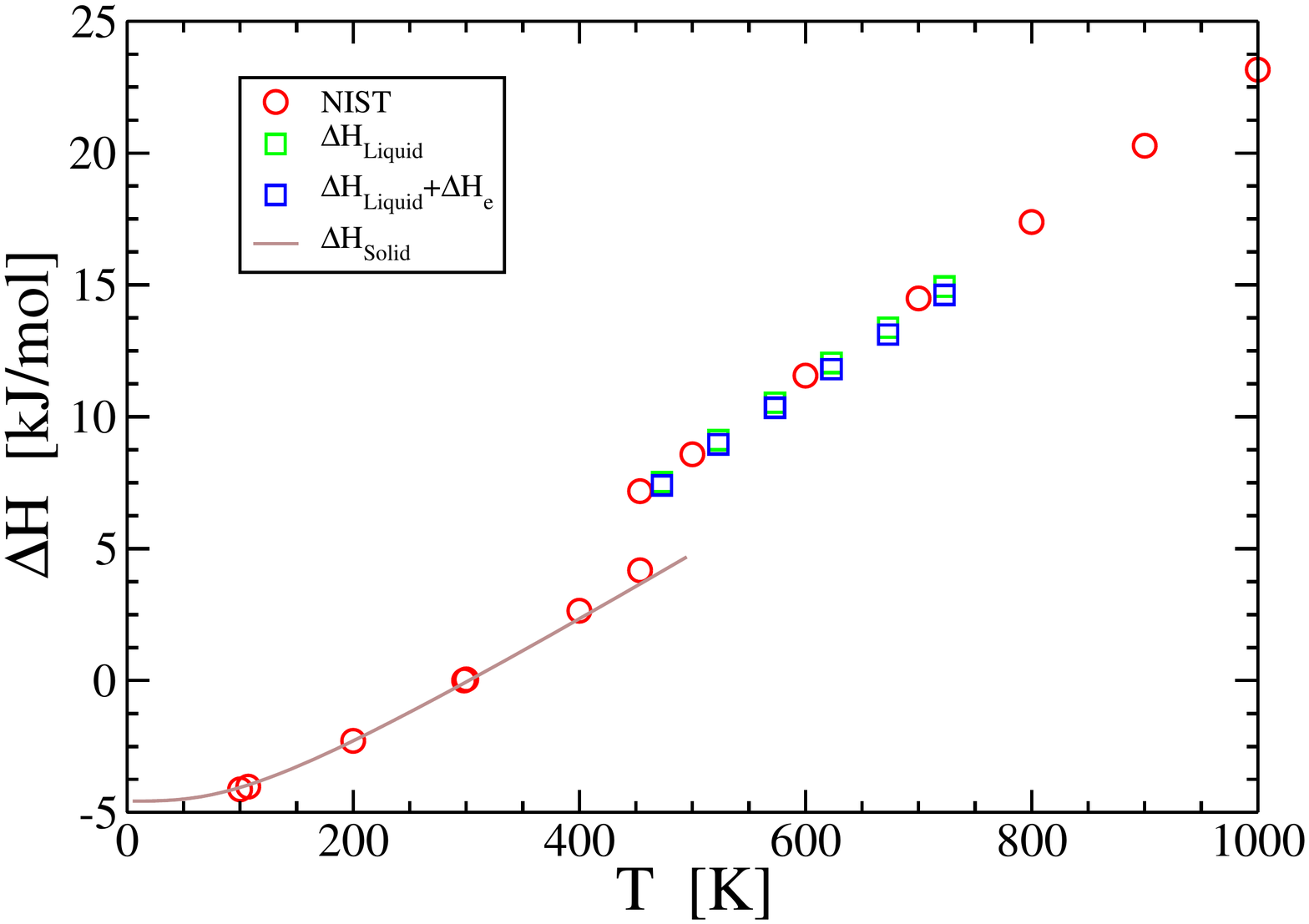}
  \includegraphics[width=.3\textwidth, height=.2\textwidth]{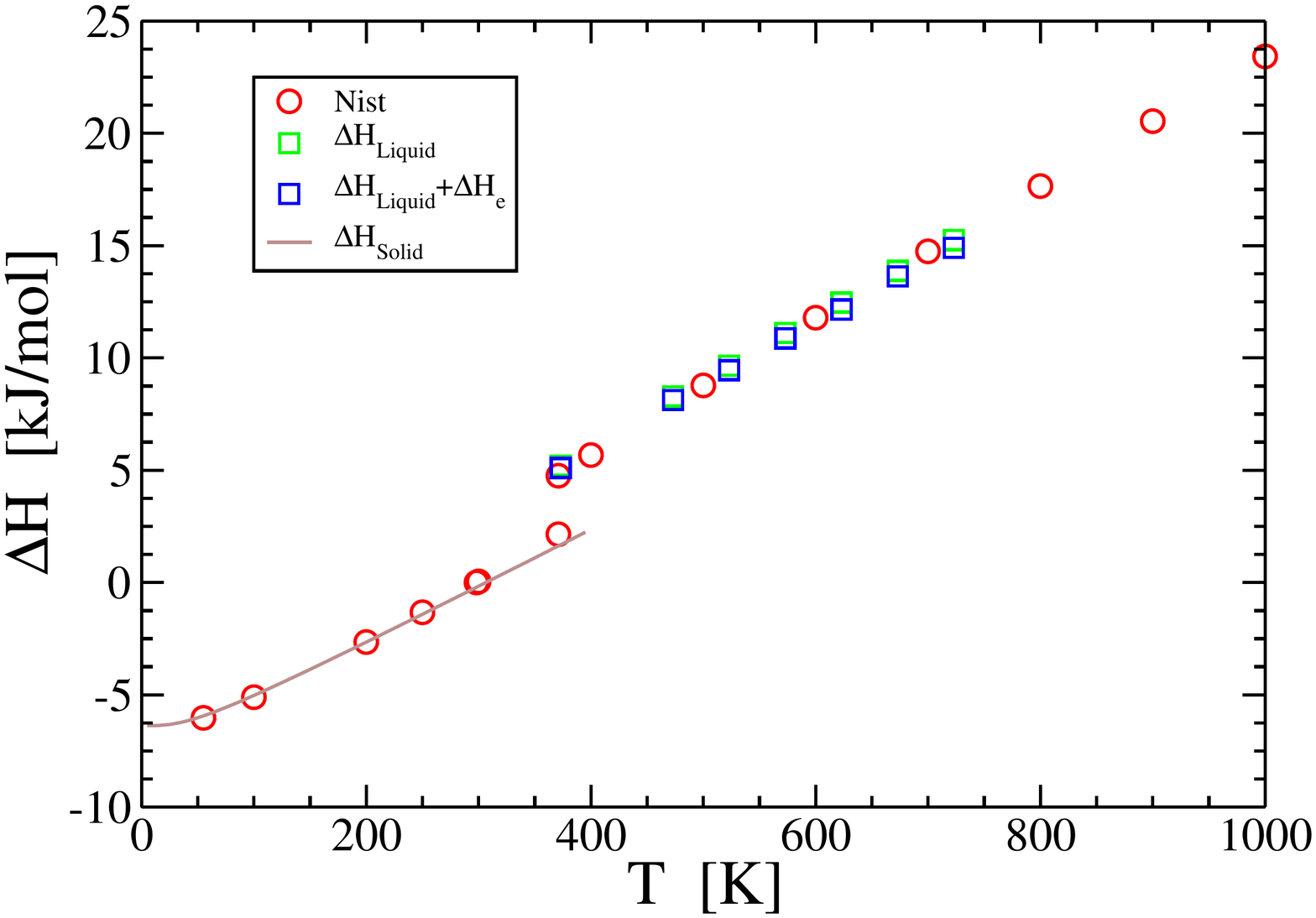}
  \includegraphics[width=.3\textwidth, height=.2\textwidth]{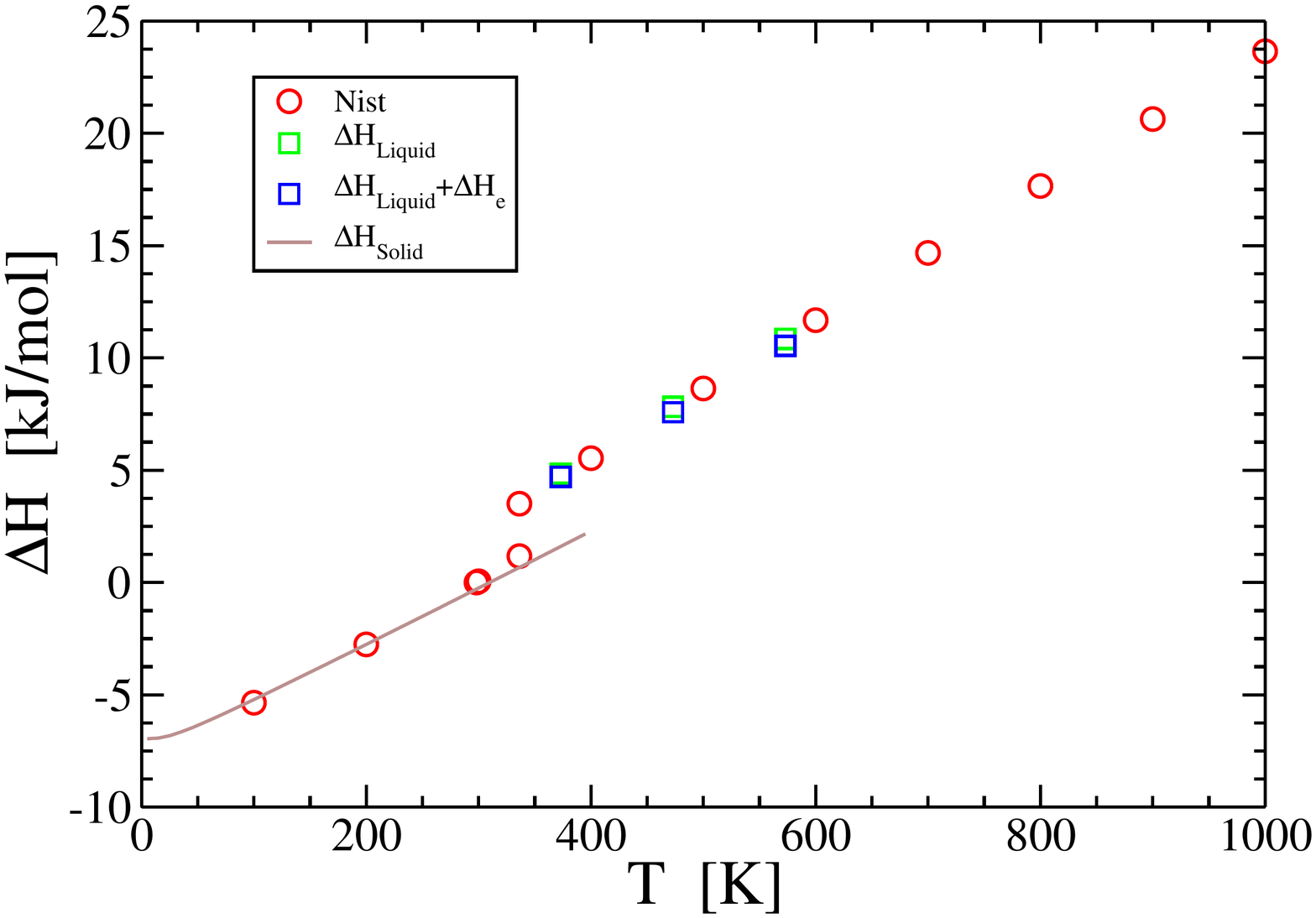}
  \caption{Comparison between calculated and experimental entropies
    (top row) and enthalpies (bottom row) for Li, Na and K (from left
    to the right). Experimental values labeled NIST come from the
    NIST-JANAF standard reference database~\cite{JANAF}, while
    experimental entropies at low temperatures are integrated from the
    heat capacities $C_p$ tabulated by
    Hultgren~\cite{alma991013058149704436}. Three liquid state entropy
    approximations are represented: the ideal gas entropy
    $S_{\rm Ideal}$, and the sum of the one-body and two-body
    entropies $S_1+S_2$ with and without electronic entropy
    $S_e$. Solid state entropies $S_{\rm Solid}$ are derived from
    phonopy.}
  \label{fig:result-entropy}
\end{figure*}
The calculated entropies and enthalpies of pure elemental Li, Na and K in their solid and liquid states are plotted in Fig.~\ref{fig:result-entropy} and compared to experimental values from the NIST-JANAF tables~\cite{JANAF}. After shifting our reference point for enthalpy to set $\Delta H=0$ at $T$=293.15K, the calculated enthalpies are in excellent agreement with experiment across both solid and liquid states, with deviations of 1 kJ/mol or less. Solid state entropies are also in excellent agreement with experiment at low temperatures but show a slight deficit of less than 1 J/mol/K just below the melting temperatures. We compare three different models for the liquid state entropy. The ideal gas model $S_{\rm Ideal}$ substantially overestimates the entropy, while the single-body entropy $S_1$ yields an improvement, and the two-body correction $S_2$ brings the value close to experiment, but slightly below. Finally, a small contribution from the electronic entropy provides an excellent match to experiment at temperatures above melting. A small deficit remains in the liquid entropy close to the melting point that is presumably due to three- and four-body correlations~\cite{Widom2019}.

\subsection{Binary K-Na}
\label{sec:K-Na}

\begin{figure}
  \includegraphics[width=.45\textwidth]{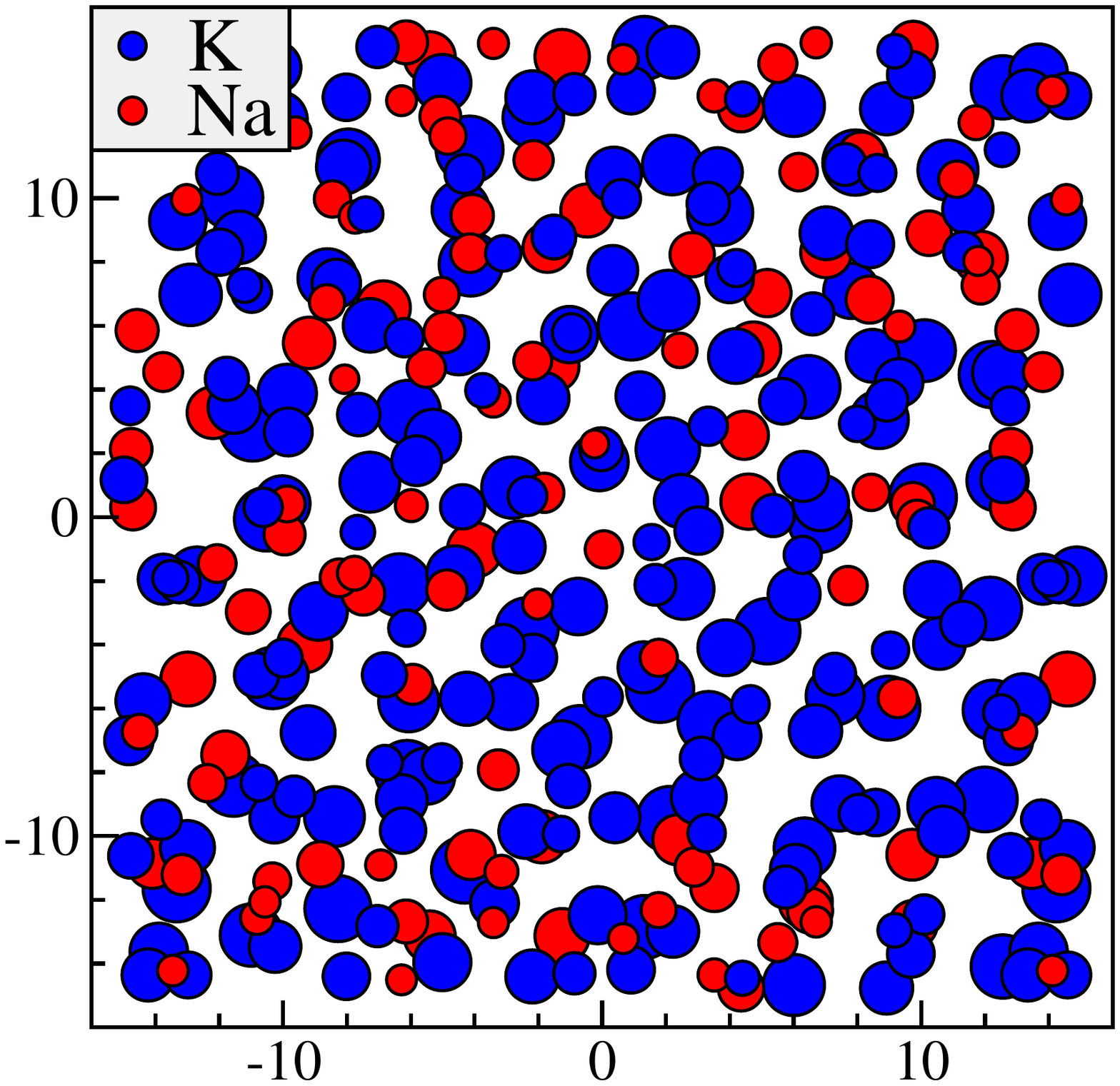}
  \includegraphics[width=.45\textwidth]{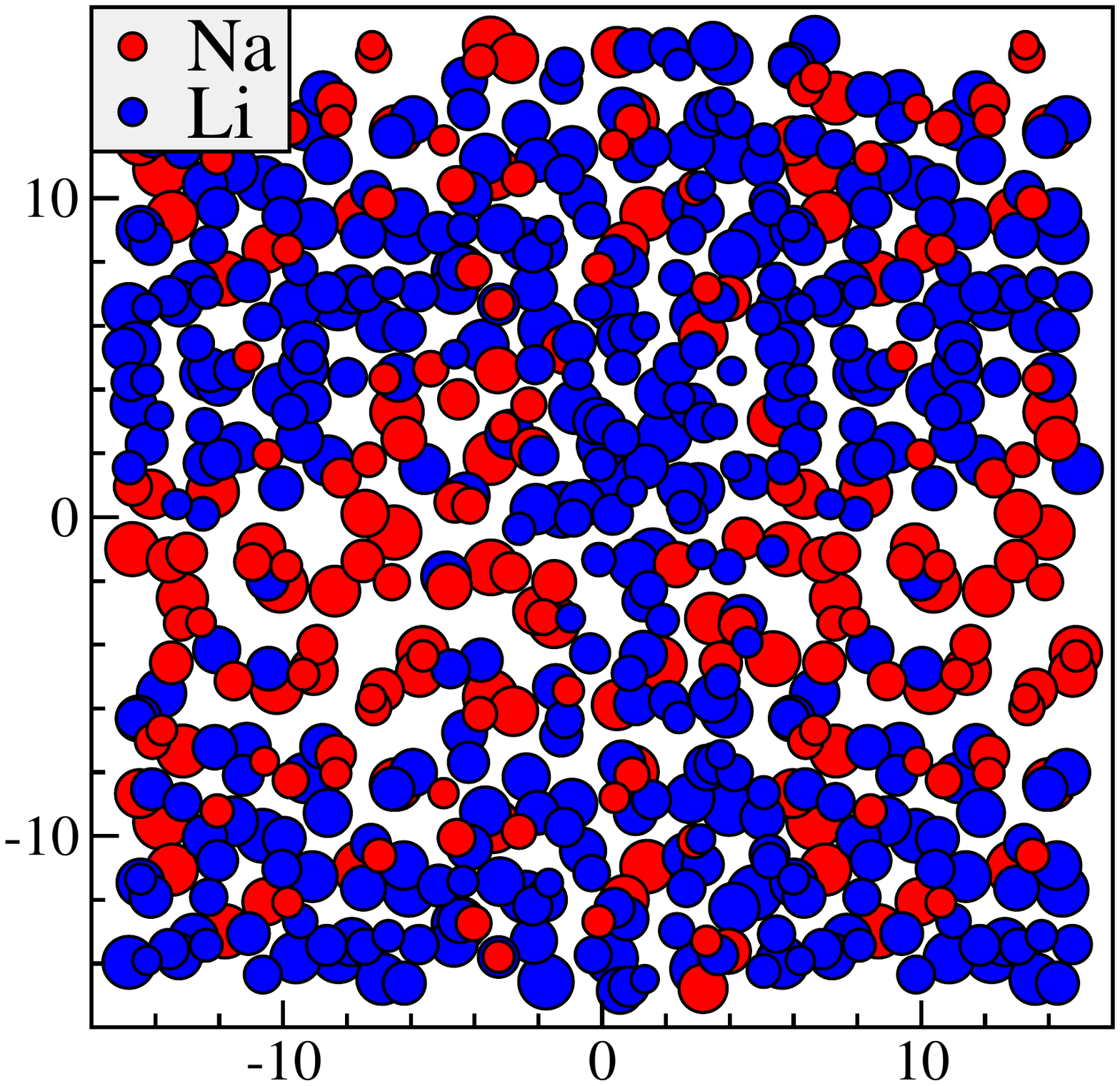}
  \caption{\label{fig:snapshots}Snapshots of typical simulated configurations for (a) K$_2$Na and (b) Li$_2$Na at T=473K. Positions are plotted from back to front, with diameters indicating depth.}
\end{figure}

\begin{figure}
  \includegraphics[width=.45\textwidth]{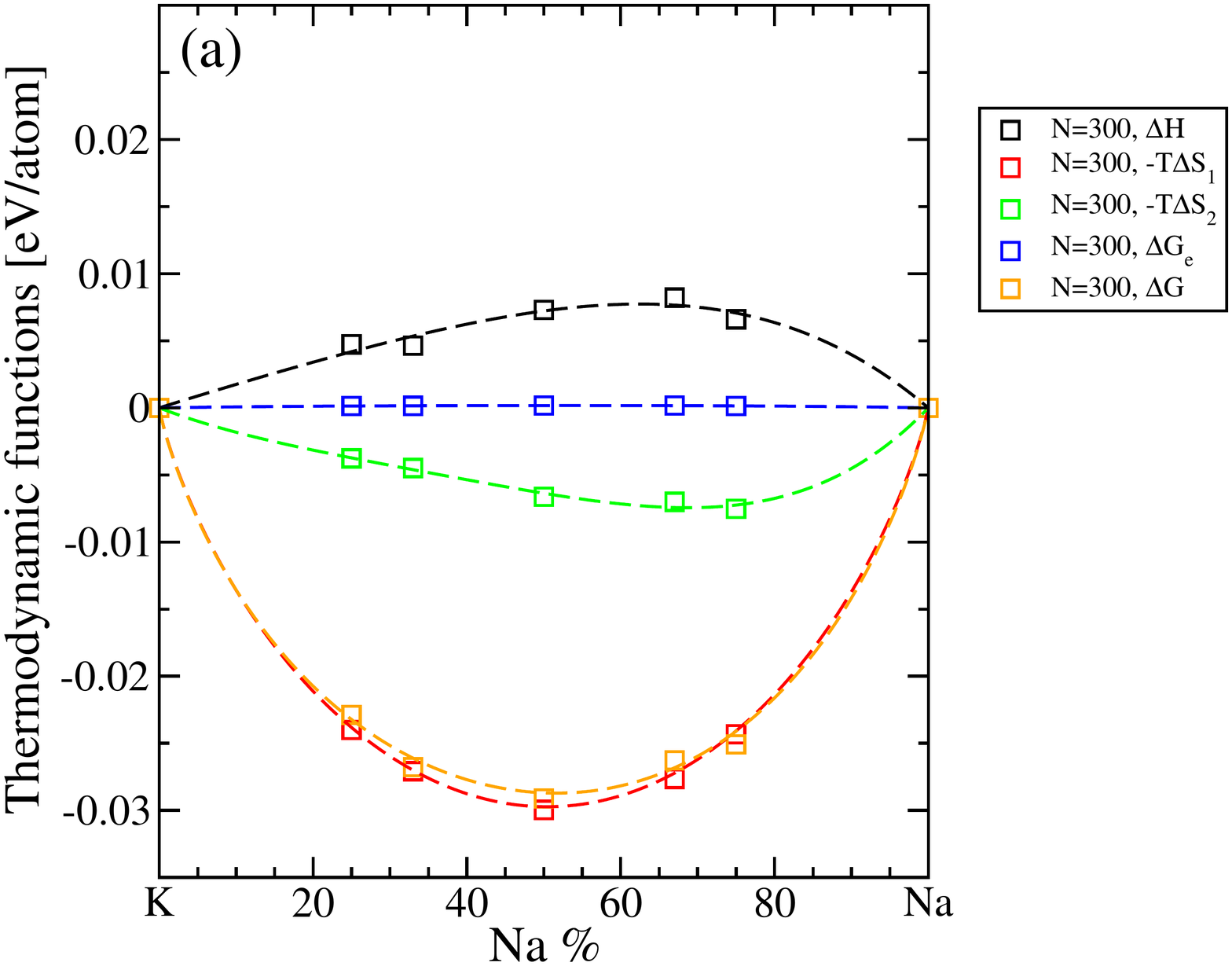}
  \includegraphics[width=.45\textwidth]{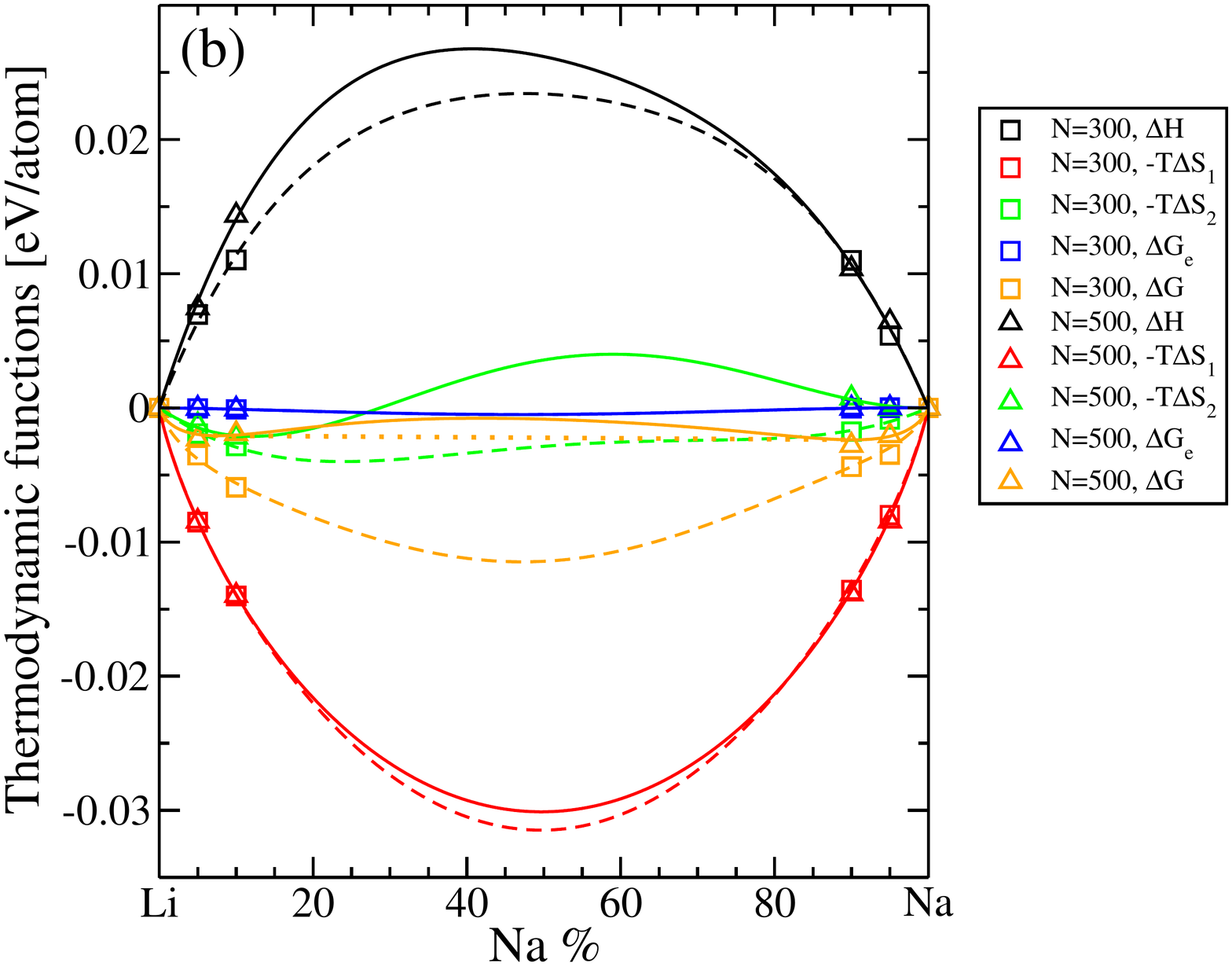}
  \includegraphics[width=.45\textwidth]{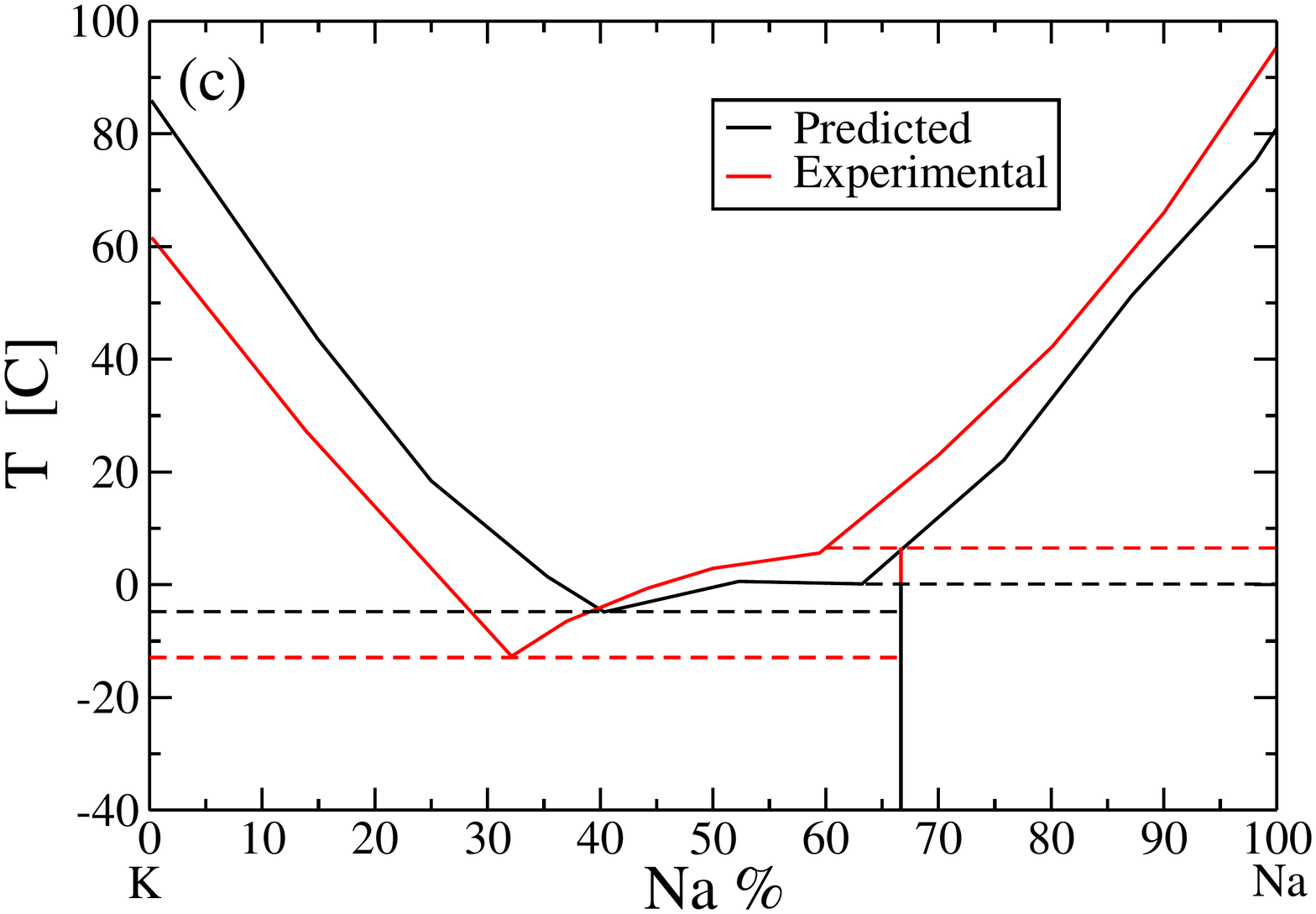}
  \includegraphics[width=.45\textwidth]{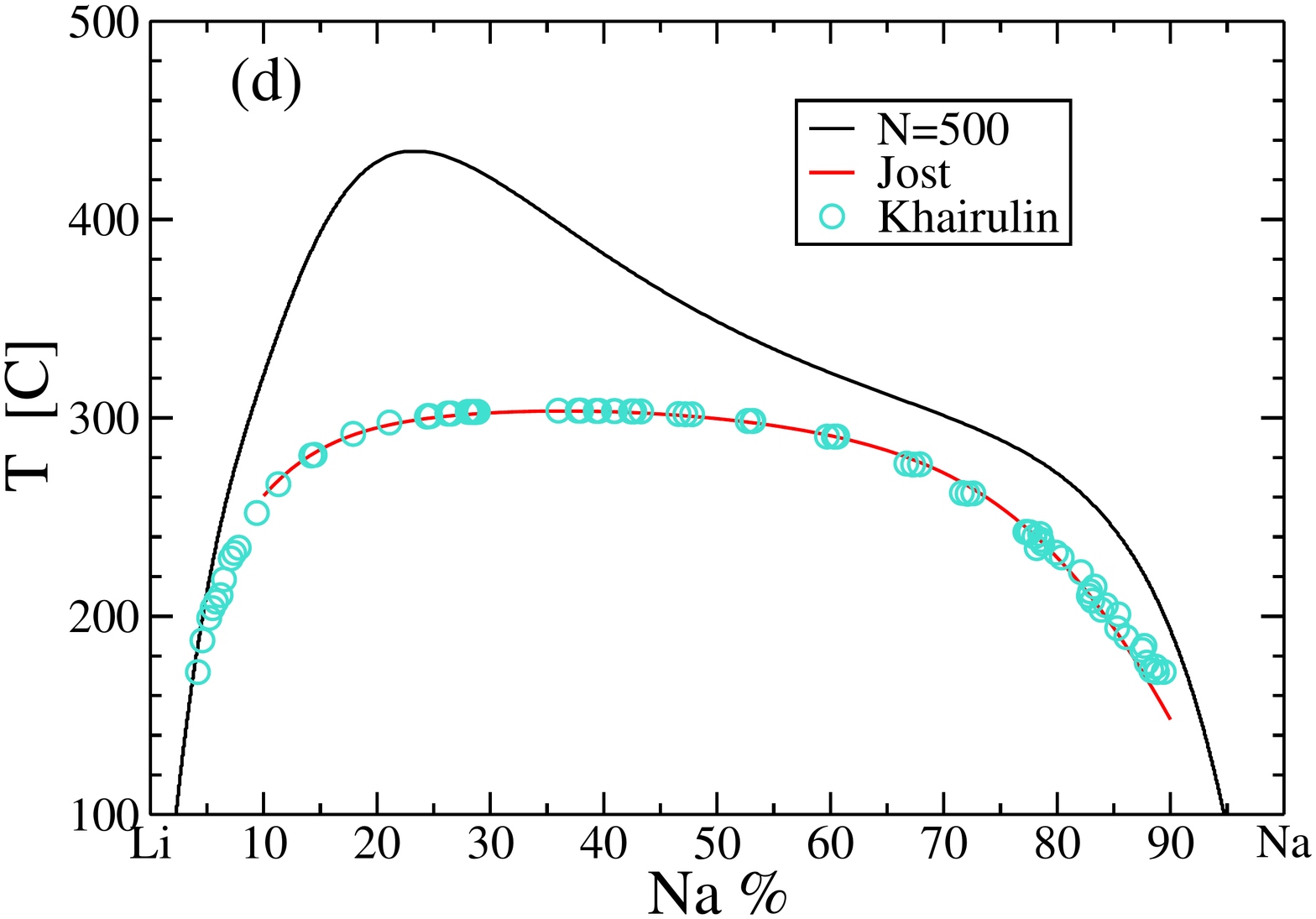}
  \caption{\label{fig:thermo} Top row: Thermodynamic functions of (a) K-Na and (b) Li-Na at $T$=473K, relative to pure elements. $S_1$ and $S_2$ are the one- and two-body contributions to the entropy. Data points are individual simulations, while curves are fits to Eq.~(\ref{eq:fit}) (dashed are from $N$=300 atoms, solid from $N$=500). The orange dotted line in (b) shows the convex hull of $G(x)$. Bottom row: Predicted and experimental phase diagrams~\cite{Bale1990a,Jost1994,Khairulin2019} of (c) K-Na and (d) Li-Na.}
\end{figure}

As shown in our simulation snapshot of K$_2$Na at T=473K in Fig.~\ref{fig:snapshots}, K and Na atoms are uniformly distributed, indicating a homogeneous liquid state. We confirm a stable mixture of liquid K-Na alloy at T=473K and $x=33\%$. The stability of the liquid alloy is confirmed by our calculated enthalpies, entropies and Gibbs free energies as plotted in Fig.~\ref{fig:thermo}. The internal energy of compound is positive, which could indicate phase separation, but the amplitude of its peak is substantially below the amplitude of the peak in the Li-Na system, and far below the magnitude of $TS$. Values of $S_1$ contain the density $\rho$, which varies monotonically and smoothly with composition, and also contains the entropy of mixing $S_{\rm Mix}$. Hence $-T\Delta S_1$ is negative and strongly convex.  Note that although $S_2$ is negative-definite, we find that $-T\Delta S_2$, defined relative to the pure elements, is  negative and seemingly is also convex. As a result, the total Gibbs free energy is dominated by entropy and is convex over all compositions, resulting at continuous miscibility of K and Na at 473K. We would expect phase separation below T=189K based on extrapolation of $\Delta G$ to low temperatures, but this is preempted by the eutectic transition to the solid phases, as we now discuss.

We compare the free energy of the liquid phase with its competing solid phases in order to predict the composition- and temperature-dependent K$_{1-x}$-Na$_x$ phase diagram as shown in Fig.~\ref{fig:thermo}~(d). K-Na has three known low-temperature phases--elemental K ($x=0$), elemental Na ($x=1$) and the KNa$_2$ Laves phase ($x=2/3$).  In the experimental phase diagram, a deep eutectic transition occurs at $x_E\approx 33\%$ and $T_E\approx 260k$, where the K-Na alloy exists in the liquid state at temperatures below the melting points of elemental K and Na. In our predicted phase diagram, a eutectic transition is found near $T_E=268K$ and $x_E=40.3\%$, not very far from the experimental position.

The deviation of these two transition point might be an effect of the systematic error in prediction via DFT or it might be due to our approximations for the entropy. To understand which is most responsible, we compare the predicted melting points ({\em i.e.} the temperatures at which solid and liquid free energies cross) with experiment as seen in Fig.~\ref{fig:thermo}~(d). Our calculated melting temperatures of elemental K and Na are  approximately $359K$ and $355K$, respectively, which differ somewhat from the experimental values of $336K$ and $370K$. An alternative approach to calculating melting temperatures via DFT using interface pinning predicts the melting point of elemental Na to be $T_m\approx 354K$\cite{Pedersen2013} which is very close to our predicted 355K.  This agreement between our approach and interface pinning suggests the discrepancy of both predictions compared with experiment may lie primarily within DFT.

\subsection{Li-Na}
\label{sec:Li-Na}
We present a snapshot of a 300-atom Li$_2$Na system at T=$473$K in Fig.~\ref{fig:snapshots}(b). The snapshot shows clear phase separation into an Na-rich region and an Li-rich region that mutually coexist in equilibrium. The separation is also evident in the correlation functions in Fig.~\ref{fig:g_s2}(b), where the amplitude of the first two peaks of the Li-Na pair correlation function is substantially smaller than those of Li-Li and Na-Na, indicating effective Li-Na repulsion.

In order to understand how internal energies and entropies contribute to the total free energies and drive the system toward phase separation, we plot our calculated enthalpy, entropy and Gibbs free energy for Li$_{1-x}$Na$_x$ in Fig.~\ref{fig:thermo}b. The energy cost of mixing Li and Na is large compared to that of mixing K and Na (Fig.~\ref{fig:thermo}a), and similar in magnitude but opposite in sign to $-TS$. The resulting free energy lacks convexity and hence explains the separation of liquid Li and Na at low and moderately high temperatures. Specifically, $G(x)$ lies above its own convex hull over the interval from $x=0.07$ to $x=0.90$ (see dotted orange line in Fig.~\ref{fig:thermo}b). A liquid alloy in this composition range will phase separate into a mixture of those two endpoint compositions. Note that we only use data at compositions that lie within our predicted single phase regions.

Collecting data similar to that of Fig.~\ref{fig:thermo}b at higher temperatures (see Fig.~\ref{fig:Li-Na-detail} in Appendix C), we then fit the temperature evolution of the phase boundaries. Fig.~\ref{fig:thermo}d compares our predicted phase coexistence region with the experimental result labeled Jost~\cite{Jost1994} and Khairulin~\cite{Khairulin2019}. This figure is based on data from T=523K and 573K. Note that we reproduce the boundary qualitatively, including the asymmetry showing greater solubility at the Na-rich end, with the notable exception of the vicinity of the critical point. The difficulty in the vicinity of the critical point is not a surprise because the expected singularities in the thermodynamic functions cannot be represented within our polynomial form (Eq.~(\ref{eq:fit})). Similarly, the diverging correlation lengths near the critical point cannot be accommodated in our finite size simulation cells. Critical exponents of Li-Na have recently been measured and match expectations for the three-dimensional Ising universality class~\cite{Khairulin2019}.

In addition to the Li-Na phase separation, a eutectic transition (not shown) occurs at $x_{\rm Na}\approx 97\%$ and $T\approx 290K$ and a monotectic transition near the melting point of Li. Both features are also reported in the experimental phase diagram~\cite{Bale1990b}.

\section{Conclusions}
\label{sec:Conclusions}
In summary, we systematically studied thermodynamic properties of solid and liquid K-Na and Li-Na metallic alloys at finite temperature and zero pressure. The Gibbs free energies of BCC K, Li and Na, and Laves phase KNa$_2$, were calculated in the quasi-harmonic approximation as implemented in {\tt Phonopy}~\cite{phonopy,phonopy-qha}. Standard {\em ab-initio} molecular dynamics and Monte Carlo/molecular dynamics simulations modeled the liquid alloys. Absolute entropies in the liquid state were obtained as functionals of simulated densities and pair correlation functions. We note that a similar approach is possible in the solid state also~\cite{Nicholson2021,Gao2018}. At $T$=473K we observed phase separation in Li$_2$Na in contrast to phase mixing in K$_2$Na, and these observations were justified by explicit calculation of the composition-dependent absolute Gibbs free energy $G(x)$ that revealed nonconvexity in the case of Li-Na.

Extending our calculations to other temperatures, we predicted composition-temperature phase diagrams that agreed well with experiment in most respects. Specifically, we obtained a deep eutectic transition in K-Na and liquid-liquid phase separation in Li-Na. Our principal shortcoming was our inability to accurately model the critical point for Li-Na phase separation. We attribute this difficulty to the thermodynamic singularities and diverging correlation length that characterize the critical point~\cite{Khairulin2019}.

\acknowledgments
YH and MW acknowledge support of the US Department of Energy grant DE-SC0014506 for performing calculations and analyzing results. Computer time was provided at the Pittsburgh Supercomputer Center under XSEDE grant DMR160149. MCG acknowledges the support of the US Department of Energy’s Fossil Energy Crosscutting Technology Research Program through the NETL Research and Innovation Center’s Advanced Alloy Development Field Work Proposal.

\clearpage

\begin{appendix}
  \renewcommand\thefigure{\thesection.\arabic{figure}}
  \setcounter{figure}{0}  
\section{Optimization of calculation parameters}
\label{app:A}

Density functional theory calculations make numerous approximations that affect the accuracy of its predictions. Here we test choices of numerous calculational details that affect our results, seeking to achieve a balance between accuracy and computational efficiency. Because the enthalpy and entropy both depend sensitively on the density of the liquid, and experimental data on density is readily available and presumably reliable, we take the accuracy of our predicted density as a measure of accuracy overall.

\subsection{Size of simulation cell}
\label{sec:finite-size}

\begin{figure}[htpb]
  \label{fig:optN}
  \includegraphics[width=.4\textwidth]{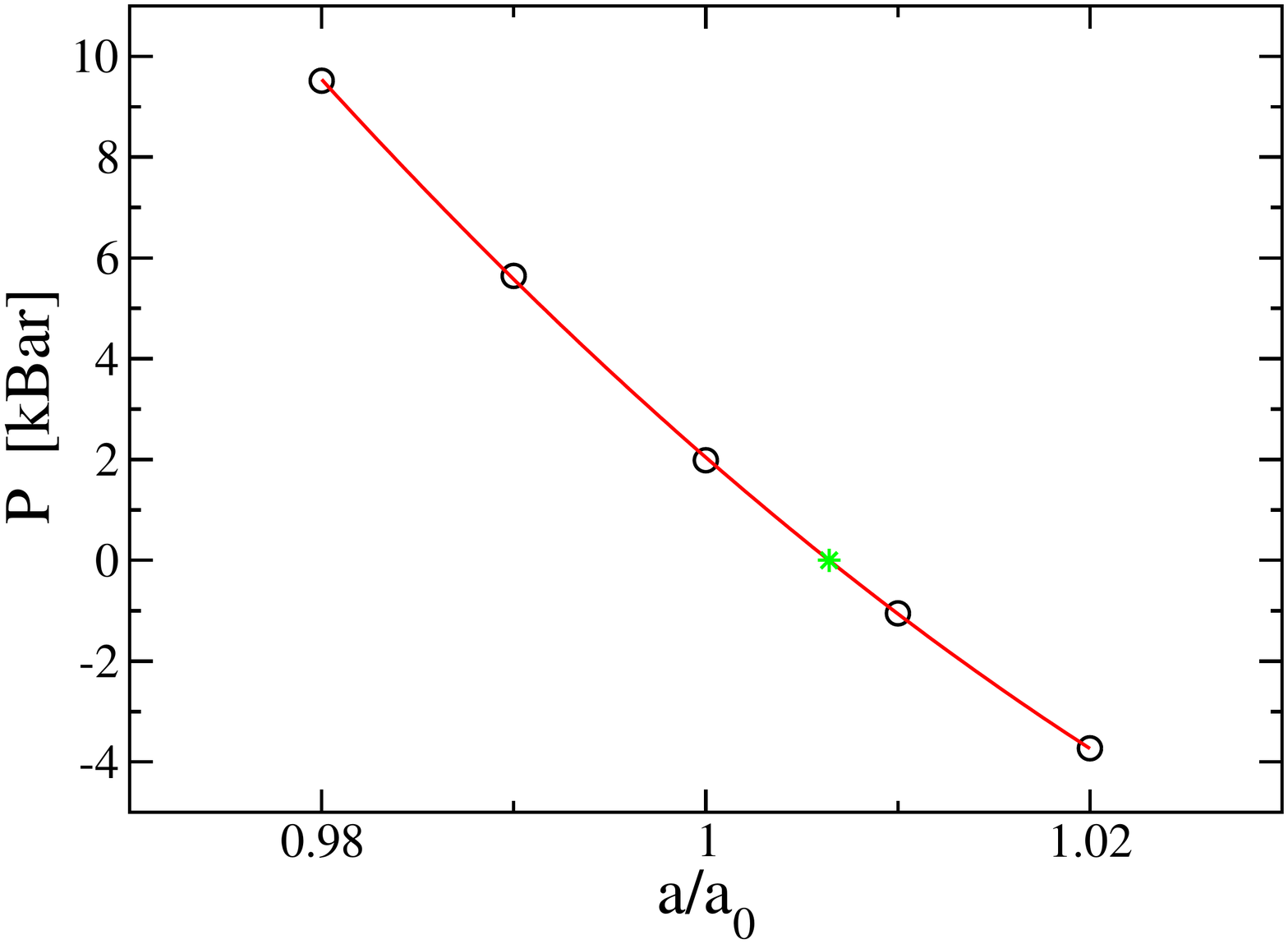}
  \includegraphics[width=.4\textwidth]{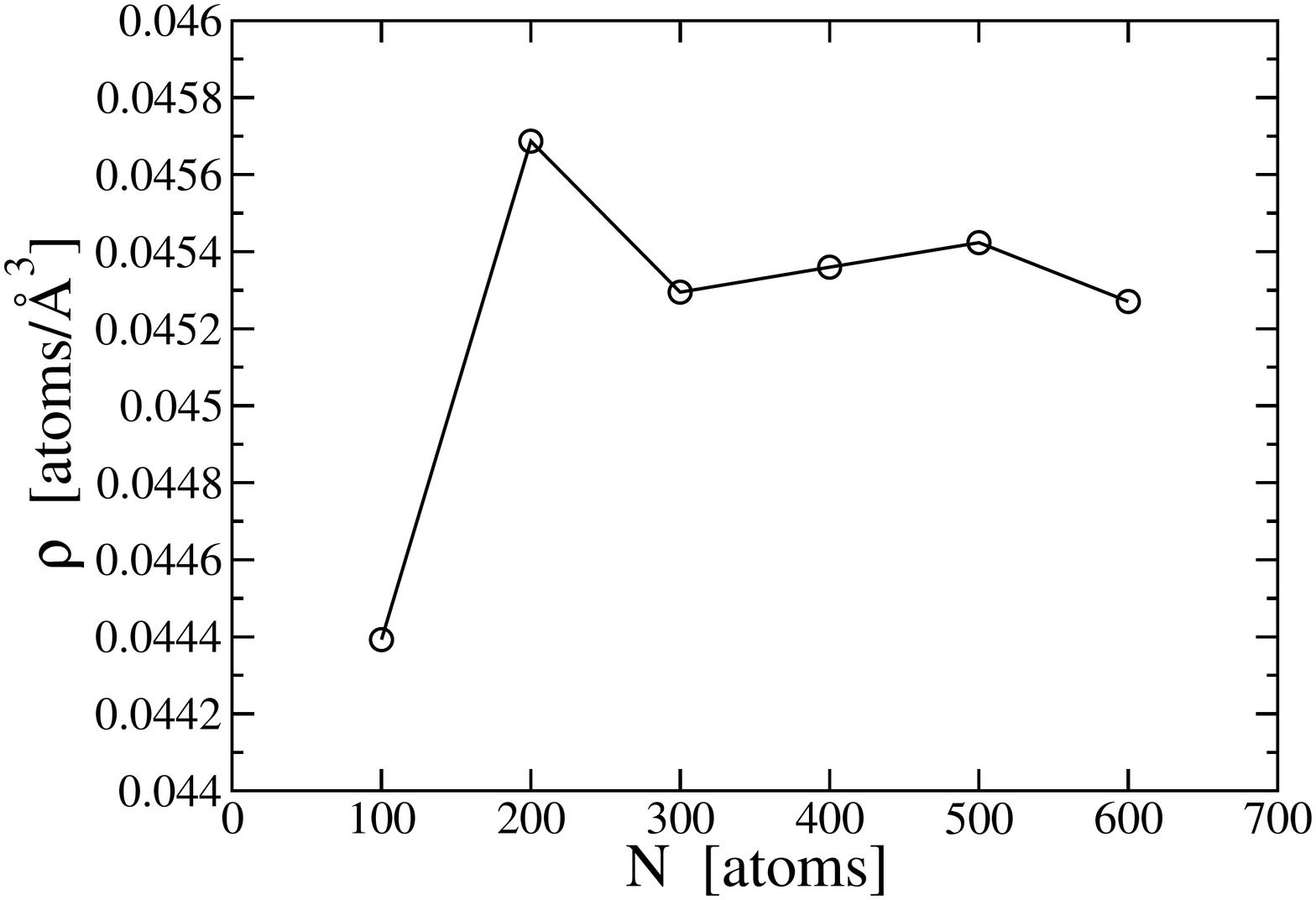}
  \caption{(a) Pressure {\em vs.} cell size $a$ for $N=300$ Li atoms at T=473K. Circles are individual runs, curve is a quadratic fit, and the star marks fitted $P=0$. (b) Variation of predicted density $rho$ with respect to number of atoms $N$.}
\end{figure}

Previously, in our study of liquid Al~\cite{Widom2019}, we observed oscillations in the predicted density as a function of the size of the simulation cell. This effect was related to the commensurability of cell lattice constant $a_0$ with the oscillation frequency of the spatial correlation functions. To assess this behavior for alkali metals we simulated elemental liquid Li using $N=100-400$ atom cells. The runs were performed using the valence-1 PAW\_PBE-type pseudopotential "Li 17Jan2003" and we employed a plane wave energy cutoff of 180 eV that exceeds the default by the recommended 30\%, and a single $k$-point. For each number of atoms we performed runs at T=473K taking several cell sizes $a$ in the vicinity of the expected optimum $a_0$. We fit the average pressure (including the kinetic component~\cite{Ganesh2009}) at each size to a quadratic and solved for the equilibrium volume at $P=0$ as illustrated in Fig.~\ref{fig:optN}a.  Oscillations die off with increasing size, and on the basis of Fig.~\ref{fig:optN}b we judge that we obtain accuracy of around 0.2\% when $N=300$. Since the ideal gas entropy varies logarithmically with respect to density, this translates into an uncertainty of around 0.002 k$_B$ in the ideal gas component of the entropy, and will be a relatively small component of our overall uncertainty.

When we simulated the Li-Na and K-Na binaries, we found that the $N=300$ atom cells exhibited spontaneous phase separation (see, e.g. main text figure 2) over wide ranges of composition, even at temperatures far above the expected critical temperature. In contrast, cells with $N=500$ did not show this behavior, and in fact revealed only localized indications of separation below the critical temperature. Thus we adopt $N=300$ atom cells for K-Na and $N=500$-atom cells for Li-Na. Further, our thermodynamic modeling is restricted to data obtained from temperature-composition combinations at which phase separation is not predicted.

\subsection{Cutoff energy}
\begin{figure}
  \label{fig:density}
  \includegraphics[width=.8\textwidth]{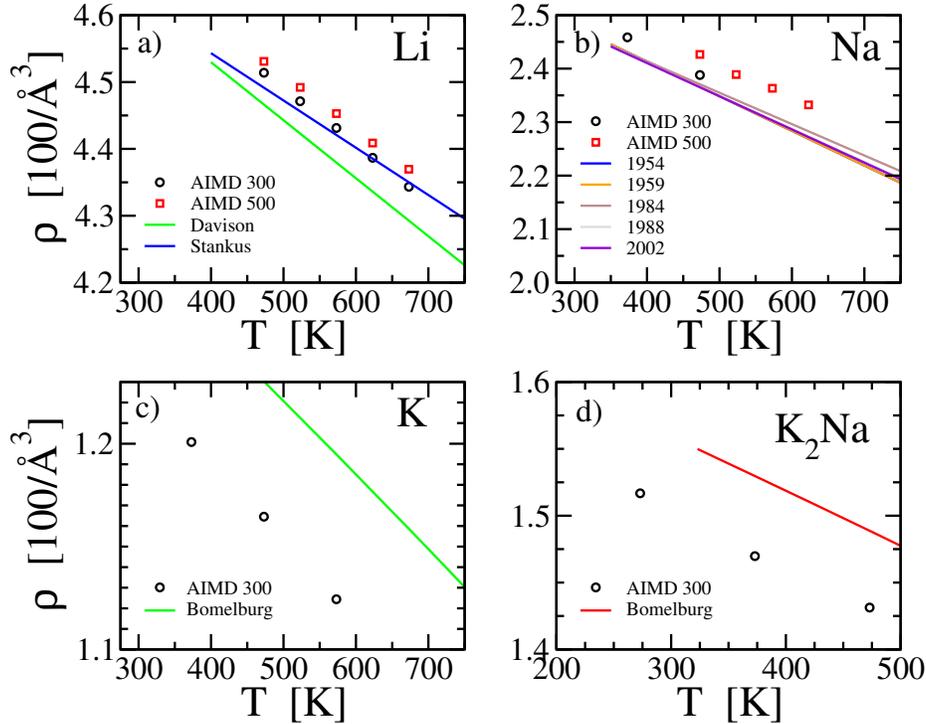}
  \caption{Comparison of AIMD predicted densities with experimental data. All AIMD runs use a single $k$-point, energy cutoff of 300 eV, and PAW potentials in the PBE gradient approximation. (a) Li with $N$=300 and 500 atoms. References are Davison~\cite{Davison1968}, Stankus~\cite{Stankus2011}. (b) Na with $N$=300 and 500 atoms. References 1954-2002 come from~\cite{Sobolev2011}. (c) K with $N$=300 atoms. (d) K$_2$Na with $N$=300 atoms. Reference Bomelburg is~\cite{Bomelburg1972}.}
\end{figure}

The incompleteness of the plane wave basis set creates systematic errors in the calculated pressure (the Pulay stress~\cite{Pulay1969}) that diminish as the plane wave energy cutoff increases.  Different elements, and even different pseudopotentials for the same element, have very different default energy cutoffs. The defaults are 140.000 eV for "Li 17Jan2003", 116.731 eV for "K\_pv 17Jan2003" and 259.561 eV for "Na\_pv 19Sep2006".  The "\_pv" subscripts indicate that $p$ semicore electrons are treated as valence. Calculational cost grows as the cube of the number of plane waves, which itself grows as the 3/2 power of the energy cutoff, leading to rapid growth of cost {\em vs.} cutoff, yet consistency of calculated energies requires that the applied energy cutoff be uniform across different compositions. Since the Na potential requires the highest energy cutoff for both the Li-Na and K-Na alloy systems, we explore the sensitivity of the density of Na to the energy cutoff. Testing values 260, 300 and 340 eV, with systems of $N=300$ atoms at T=473K, we obtained densities of 0.02948, 0.02378 and 0.02381 atoms/\AA$^3$. Thus we settle on a cutoff of 300 eV. Note the experimental value is around 0.0237 (see Fig.~\ref{fig:density})b.

Fig.~\ref{fig:density}d illustrates the composition-dependent densities for Li-Na at two temperatures and compare with experiment. Our predictions follow the experimental trends, while remaining slightly high.

Note that the predicted density of K drops with increasing energy cutoff, and converges to nearly 5\% below the experimental value at this cutoff. The situation is marginally improved if we switch to the ``K\_sv'' potential, but not sufficiently to compensate for the increased electron count.

\subsection{XC functional}
\label{sec:XC}

The choice of exchange correlation functional can lead to systematic errors in the density. We compared the local density approximation (LDA) with the PBE generalized gradient approximation and with PBEsol for a system of $N=300$ Li atoms at T=473K using an energy cutoff of 300 eV and a single $k$-point. These yielded densities of 0.04535, 0.04516 and 0.04511 atoms/\AA$^3$, for LDA, PBE and PBEsol respectively, compared with experimental values that range from 0.0444 to 0.0448.  All three overestimate the density. PBEsol proves only marginally better than PBE, and we prefer PBE because it is more widely used. To test if this is caused by neglect of the Li core electrons, we tested the valence-3 "Li\_sv 10Sep2004" potential at energy cutoff 650 eV (30\% above its default of 499.034 eV) and found density 0.0462, which is far above the experimental value. Thus we settle on the PBE functional and stick with our decision to use the valence-1 Li potential.

\subsection{Uncertainties in thermodynamic quantities}
\label{sec:Uncertainty}

Our calculations are subject to both systematic and statistical errors. Density functional theory itself relies on the approximate exchange-correlation functional, leading to systematic errors that we do not attempt to quantify, beyond noting that differing choices of functional had modest influence on the density as discussed in Sec.~\ref{sec:XC}, while DFT may lead to errors in the vicinity of 15K for the melting point of Na as discussed in Sec.~\ref{sec:K-Na}. We truncate our expansion of the entropy at the pair level, systematically omitting three- and higher-body correlations. This may be the reason that our entropy falls below experiment close to the melting point~\cite{Widom2019,Nicholson2021}, as seen in Fig.~\ref{fig:result-entropy}. We restrict our simulations to certain finite sizes, leading to errors on the density as shown in Fig.~\ref{fig:optN}, but also leading to premature phase separation in Li-Na even at high temperatures above the critical point, when the correlation length grows beyond our simulated cell size (see Fig.~\ref{fig:Li-Na-detail}).

To estimate our statistical errors, we break our runs into three segments and evaluate the standard error on the assumption of uncorrelated errors. For K$_2$Na at T=473K we find statistical errors of order 1 meV/atom in both the enthalpy $H$ and the entropy $-TS_2$. However these are anticorrelated so that the statistical error on $G$ is of order 0.4 meV/atom. These statistical errors are in general agreement with the scatter of data points around our smoothed fitting curves as seen in Figs.~\ref{fig:thermo} and~\ref{fig:Li-Na-detail}.

\section{Accelerated sampling}
\renewcommand\thefigure{\thesection}
\setcounter{figure}{0}

\begin{figure}
  \label{fig:aimd-mcmd}
  \includegraphics[width=.45\textwidth]{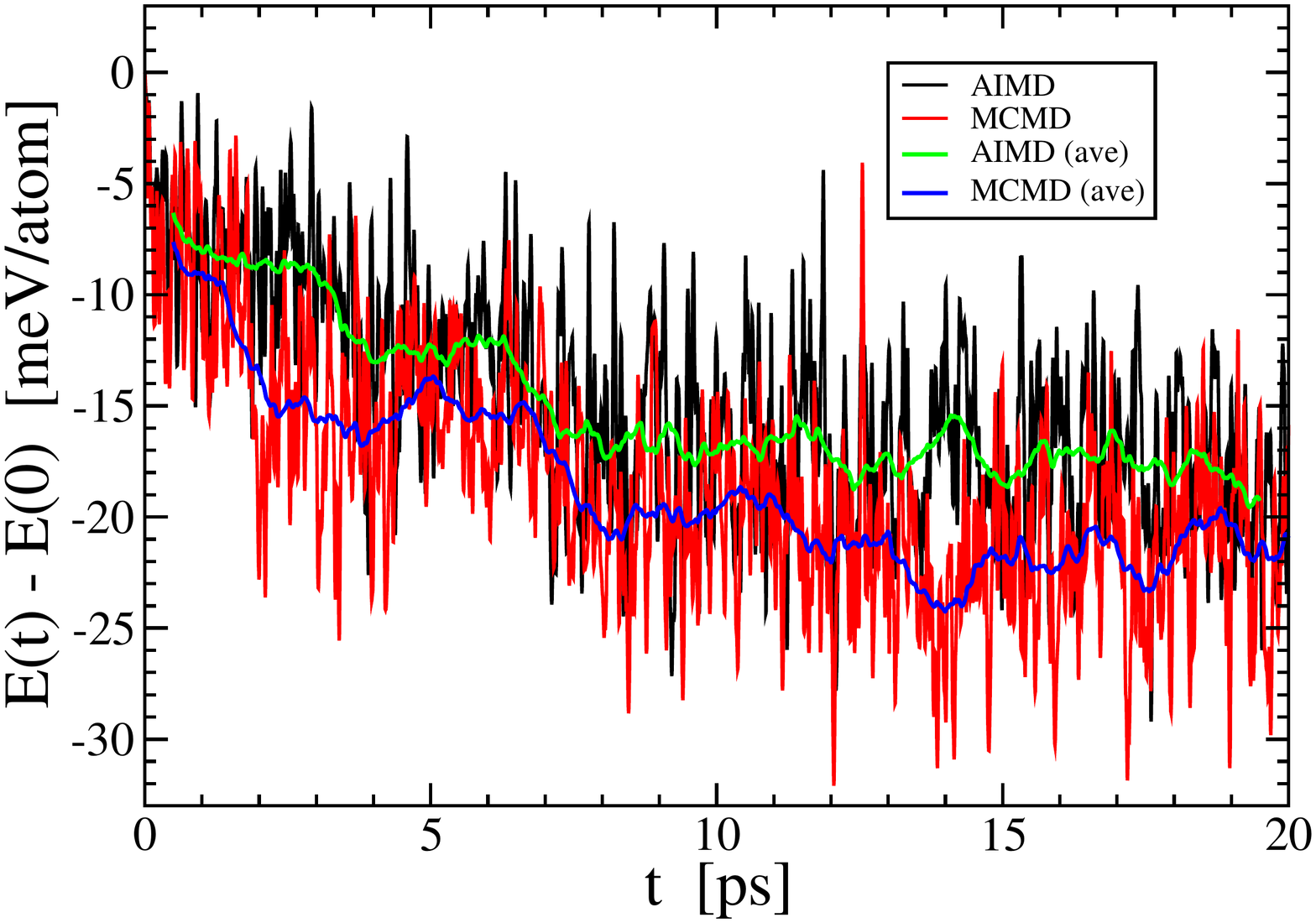}
  \includegraphics[width=.45\textwidth]{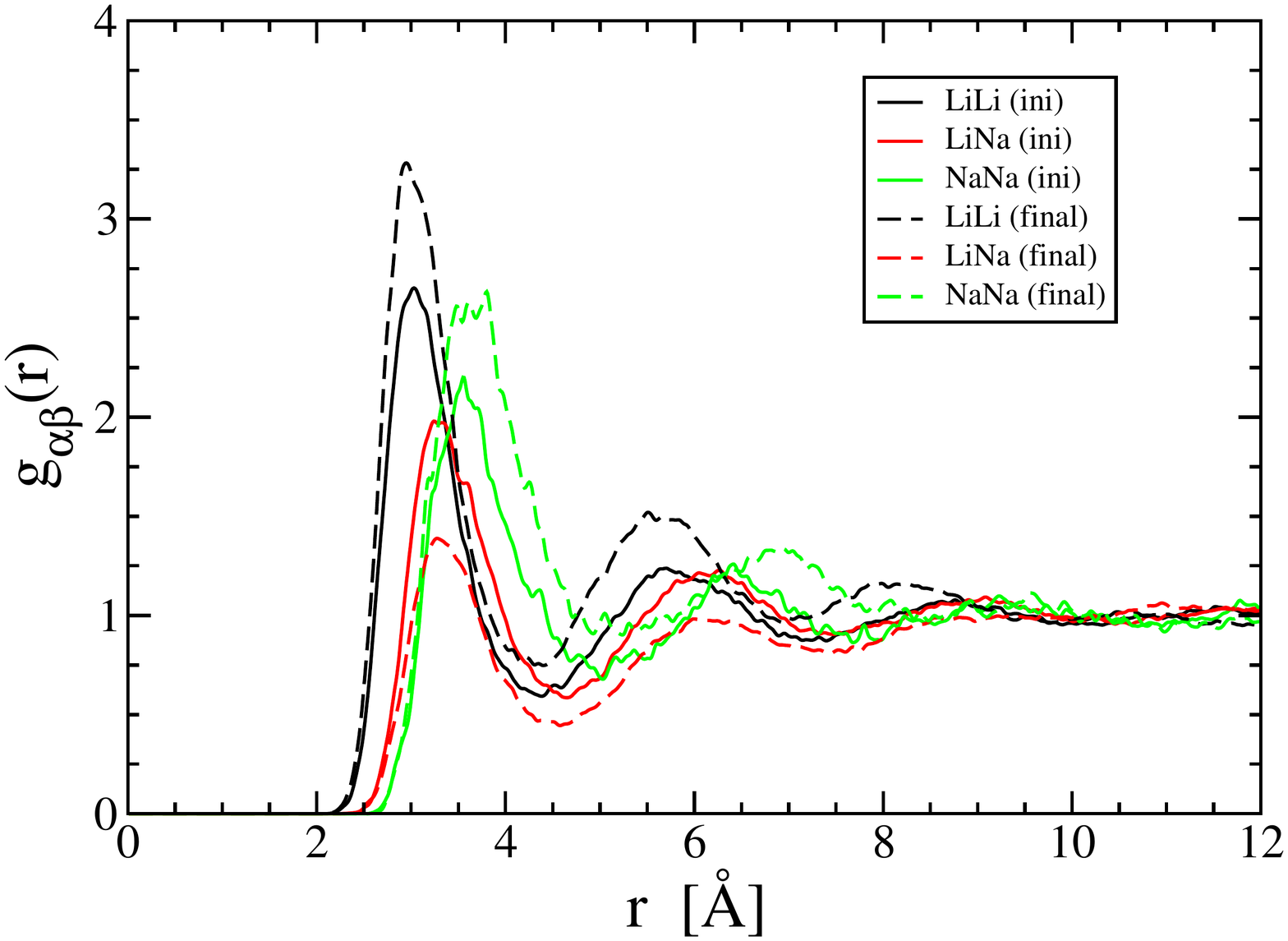}
  \caption{(a) Equilibration of total energy for AIMD compared with MCMD showing instantaneous values and running averages over intervals of 1ps. The Li$_2$Na liquid has $N$=300 atoms at T=573K. Monte Carlo swaps are attempted every 10 fs. (b) Pair distribution functions $g_{\alpha\beta}(r)$ averaged over initial and final 1ps of MCMD simulation.}
\end{figure}

Hybrid Monte Carlo/molecular dynamics (MCMD~\cite{Widom2014}) is applied in order to accelerate the sampling of the configurational ensemble. This method supplements conventional molecular dynamics, in which the structure evolves continuously through diffusion of atoms, with discrete interchanges of pairs of atoms of differing chemical species. Although the Monte Carlo steps are less important in the liquid state than in the solid state, where diffusion is nearly unachievable, we still see an improvement in equilibration time. Fig.~\ref{fig:aimd-mcmd}a graphs the evolution in total energy for a 300 atom Li$_2$Na liquid at T=573K in which atomic species have initially been randomly interchanged then briefly annealed under conventional AIMD. The energy drops more rapidly under MCMD as species swaps allow more rapid growth of clusters of like-species atoms.  Once equilibrium is achieved, around 10ps, MCMD continues to enhance the diversity of the sampled ensemble, while AIMD has not yet reached equilibrium. Fig.~\ref{fig:aimd-mcmd}b illustrates the evolution of pair correlations during the MCMD simulation, showing a drop in mixed Li-Na species pairs and corresponding growth in like species pairs.

\section{Detailed results for Li-Na}
\label{app:C}
\setcounter{figure}{0}

\begin{figure}
  \label{fig:Li-Na-detail}
  \includegraphics[width=.8\textwidth]{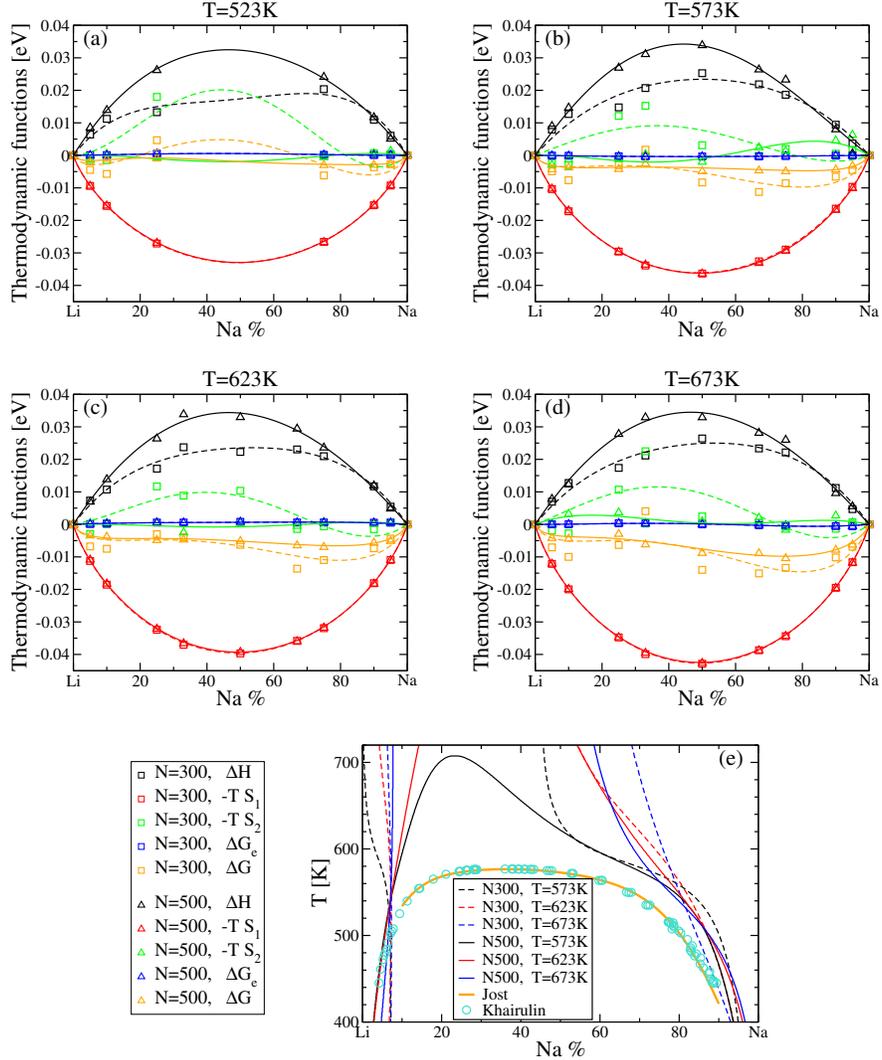}
  \caption{(a-d) Thermodynamic functions of Li-Na at temperatures 523-673K. and (e) predicted phase boundaries. Thermodynamic functions are: enthalpy $H$ (black), one-body entropy $-TS_1$ (red), two-body entropy $-TS_2$ (green), electronic free energy $G_e$ (blue), and total free energy $G$ (orange). Data points are AIMD simulation (squares for $N$=300 atoms, triangles for $N$=500). Curves (dashed for $N$=300 atoms and solid for $N$=500) are fits to Eq.~(\ref{eq:fit}). (e) Predicted phase boundaries for Li-Na including all data up to and including temperatures 573-673K from simulation sizes $N$=300 and 500 atoms. Curve labeled Jost is a model fit to experiment~\cite{Jost1994}, while explicit data poitns are taken from Khairulin~\cite{Khairulin2019}.}
\end{figure}

This section presents detailed results for the simulated thermodynamic functions of Li-Na at various temperatures and numbers of atoms, as illustrated in Fig.~\ref{fig:Li-Na-detail}a-d. Nonconvexity of the total Gibbs free energy $G(x)$ predicts phase separation at each temperature, though the uncertainties on our fits to Eq.~(\ref{eq:fit}) are sufficient to render those predicts doubtful especially at the higher temperatures. Consequently, our predicted phase boundaries place the critical points for phase mixing far above the experimental value (see Fig.~\ref{fig:Li-Na-detail}), especially for $N$=300 atoms, where visible phase separation (see Fig~\ref{fig:snapshots}) persists to the highest temperatures. The experimentally observed asymmetry in the coexistence curve is respected in every predicted phase boundary.

\end{appendix}

\newpage

\center{\Large REFERENCES}
\bibliography{refs}

\newpage

\center{\Large SUPPLEMENTAL DATA}

The following are individual calculated data points for Fig. 4 and Fig. C. Units are eV/atom.

\begin{table}[htpb]
  \centering
  \begin{tabular}{|l|l|l|l|l|l|}
    \hline
    Na \% &  G & H & TS$_1$ & TS$_2$\\
    \hline
    Solid Na &1.597&1.732&0.135&\\
    100  &  -1.56909   & -1.22519  &   0.43124       &-0.08735\\
    95  &  -1.59423   & -1.24799   &  0.43477      &-0.08854\\
    90 &   -1.61799  &  -1.27327   &  0.43531    &-0.09059\\
    10 &   -1.98617  &  -1.73692    & 0.35697   & -0.10772\\
    5 &   -2.00970  &  -1.77305     &0.34654  &  -0.10989\\
    0&    -2.03048 &   -1.80972     &0.33318&    -0.11242\\
    Solid Li&4.1581&4.1873&0.0291&\\
    \hline
  \end{tabular}
  \caption{LiNa T=473K}
\end{table}

\begin{table}[htpb]
  \centering
  \begin{tabular}{|l|l|l|l|l|l|}
    \hline
    Na \% &  G & H & TS$_1$ & TS$_2$\\
    \hline
    100 &    -1.61442 &   -1.21522 &     0.48433 &   -0.08512\\
    95&    -1.63979 &   -1.23936   &     0.48813  &  -0.08770\\
    90&    -1.66253 &   -1.26201   &     0.48872  &  -0.08820\\
    75&    -1.73033 &   -1.33730   &     0.48384  &  -0.09081\\
    25&    -1.95403 &   -1.62757   &     0.42968  &  -0.10323\\
    10&    -2.02135 &   -1.72756   &     0.40193  &  -0.10814\\
    5&    -2.04492  &  -1.76217    &     0.39042  &  -0.10767\\
    0&    -2.06524 &   -1.79992    &     0.37558  &  -0.11026\\
    \hline
  \end{tabular}
  \caption{LiNa T=523K}
\end{table}

\begin{table}[htpb]
  \centering
  \begin{tabular}{|l|l|l|l|l|l|}
    \hline
    Na \% &  G & H & TS$_1$ & TS$_2$\\
    \hline
    100&-1.66200&-1.20462& 0.53792&-0.08054\\
    95&-1.68706&-1.23339& 0.54206&-0.08839\\
    90&-1.71002&-1.25541& 0.54279&-0.08818\\
    75&-1.77603&-1.32823& 0.53745&-0.08965\\
    67&-1.81062&-1.37213& 0.53172&-0.09323\\
    50&-1.88516&-1.46453& 0.51469&-0.09406\\
    33&-1.95713&-1.56716& 0.49168&-0.10171\\
    25&-1.99385&-1.61834& 0.47818&-0.10267\\
    10&-2.05910&-1.71873& 0.44758&-0.10721\\
    5&-2.08157&-1.75379& 0.43492&-0.10714\\
    0&-2.09920&-1.79213& 0.41868&-0.11161\\
    \hline
  \end{tabular}
  \caption{LiNa T=573K}
\end{table}

\begin{table}[htpb]
  \centering
  \begin{tabular}{|l|l|l|l|l|l|}
    \hline
    Na \% &  G & H & TS$_1$ & TS$_2$\\
    \hline
    100&-1.70935&-1.19911& 0.59231&-0.08207\\
    95&-1.73638&-1.22348& 0.59687&-0.08397\\
    90&-1.75820&-1.24558& 0.59759&-0.08497\\
    75&-1.82361&-1.32153& 0.59164&-0.08956\\
    67&-1.85723&-1.36242& 0.58545&-0.09065\\
    50&-1.92863&-1.45817& 0.56667&-0.09622\\
    33&-2.00051&-1.55657& 0.54193&-0.09800\\
    25&-2.03480&-1.61075& 0.52708&-0.10302\\
    10&-2.09744&-1.71088& 0.49372&-0.10716\\
    5&-2.11826&-1.74695& 0.47986&-0.10855\\
    0&-2.13604&-1.78312& 0.46248&-0.10957\\
    \hline
  \end{tabular}
  \caption{LiNa T=623K}
\end{table}

\begin{table}[htpb]
  \centering
  \begin{tabular}{|l|l|l|l|l|l|}
    \hline
    Na \% &  G & H & TS$_1$ & TS$_2$\\
    \hline
    100&-1.55533&-1.20875& 0.43189&-0.08532\\
    75&-1.52885&-1.13164& 0.47165&-0.07444\\
    67&-1.51351&-1.10747& 0.47992&-0.07387\\
    50&-1.48121&-1.06044& 0.49273&-0.07195\\
    33&-1.44368&-1.01515& 0.50034&-0.07181\\
    25&-1.42325&-0.99250& 0.50220&-0.07145\\
    0&-1.34851&-0.92672& 0.49363&-0.07184\\
    \hline
  \end{tabular}
  \caption{KNa T=473K}
\end{table}

\begin{table}[htpb]
  \centering
  \begin{tabular}{|l|l|l|l|l|l|}
    \hline
    Na \% &  G & H & TS$_1$ & TS$_2$\\
    \hline
    Solid Na &1.597&1.732&0.135&\\
    100&-1.46824&-1.22556& 0.32820&-0.08551\\
    75&-1.43376&-1.15177& 0.35943&-0.07744\\
    67&-1.41816&-1.12605& 0.36608&-0.07397\\
    50&-1.38068&-1.07632& 0.37615&-0.07179\\
    33&-1.34256&-1.03029& 0.38226&-0.06998\\
    25&-1.32153&-1.00909& 0.38368&-0.07125\\
    0&-1.24744&-0.94316& 0.37683&-0.07255\\
    solid K& 1.3702& 1.6537&0.28349&\\
    \hline
  \end{tabular}
  \caption{KNa T=373K}
\end{table}

\begin{table}[htpb]
  \centering
  \begin{tabular}{|l|l|l|l|l|l|}
    \hline
    Na \% &  G & H & TS$_1$ & TS$_2$\\
    \hline
    Solid Na &-0.90524&-0.7889&0.11634&\\
    100&-1.38538&-1.24677& 0.22859&-0.08998\\
    75&-1.34580&-1.17233& 0.25133&-0.07786\\
    67&-1.32771&-1.14602& 0.25624&-0.07455\\
    50&-1.28798&-1.09931& 0.26352&-0.07485\\
    33&-1.24941&-1.05165& 0.26802&-0.07026\\
    25&-1.22634&-1.02938& 0.26904&-0.07208\\
    0&-1.15342&-0.96322& 0.26400&-0.07380\\
    solid K&-1.1281 & -0.87633&0.25172&\\
    \hline
  \end{tabular}
  \caption{KNa T=273K}
\end{table}

\end{document}